# Low temperature synthesis of heterostructures of transition metal dichalcogenide alloys ($W_xMo_{1-x}S_2$) and graphene with superior catalytic performance for hydrogen evolution

Yu Lei[1†], Srimanta Pakhira[2,3,4,5†], Kazunori Fujisawa[6], Xuyang Wang[7], Oluwagbenga Oare Iyiola[2,3,4,5], Néstor Perea López[6,8], Ana Laura Elías[6,8], Lakshmy Pulickal Rajukumar[1,8], Chanjing Zhou[1], Bernd Kabius[1], Nasim Alem[1], Morinobu Endo[9], Ruitao Lv[7*], Jose L. Mendoza-Cortes[2,3,4,5*], and Mauricio Terrones[1,6,8,9,10*]

[1]Department of Materials Science and Engineering & Materials Research Institute, The Pennsylvania State University, University Park, Pennsylvania 16802, United States; [2]Department of Chemical & Biomedical Engineering, FAMU-FSU College of Engineering, Florida State University (FSU), Tallahassee, Florida, 32310, United States; [3]Department of Scientific Computing, 400 Dirac Science Library, FSU, Tallahassee, Florida, 32304, United States; [4]Materials Science and Engineering Program, High Performance Materials Institute; FSU, Tallahassee, Florida, 32310; [5]Condensed Matter Theory, National High Magnetic Field Laboratory (NHMFL), FSU, Tallahassee, Florida, 32310, United States; [6]Department of Physics, The Pennsylvania State University, University Park, Pennsylvania 16802, United States; [7]Key Laboratory of Advanced Materials (MOE), School of Materials Science and Engineering, Tsinghua University, Beijing, 100084, China; [8]Center for 2-Dimensional and

Layered Materials, The Pennsylvania State University, University Park, Pennsylvania 16802, United States; [9]Institute of Carbon Science and Technology, Shinshu University, Wakasato 4-17-1, Nagano, 380-8553, Japan; and [10]Department of Chemistry, The Pennsylvania State University, University Park, Pennsylvania 16802, United States. [†]The authors Y.L. and S.P. contributed equally in this article. Correspondence and requests for materials should be addressed to R.L, J.L.M-C. or M.T. (email: lvruitao@tsinghua.edu.cn, jmendozacortes@fsu.edu, or mut11@psu.edu).

ABSTRACT

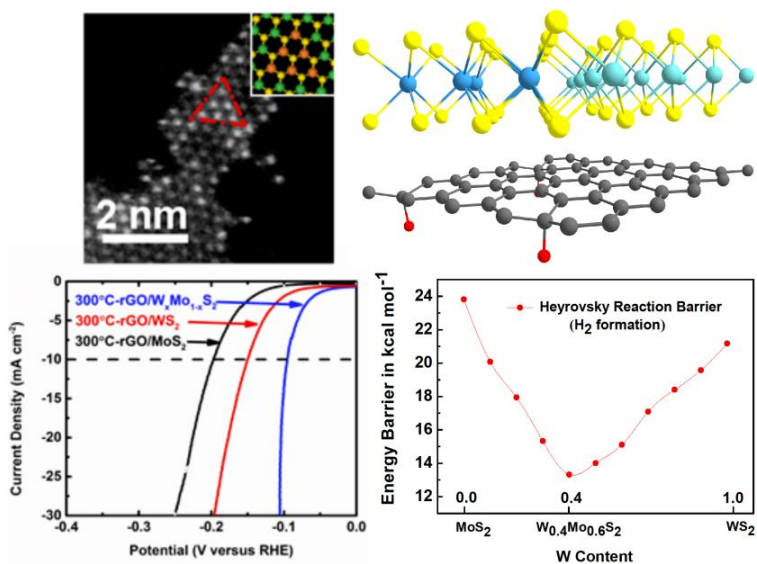

Large-area (~cm$^2$) films of vertical heterostructures formed by alternating graphene and transition-metal dichalcogenide (TMD) alloys are obtained by wet chemical routes followed by a thermal treatment at low temperature. In particular, we synthesized stacked graphene and $W_xMo_{1-x}S_2$ alloy phases that were used as hydrogen evolution catalysts. We observed a Tafel slope of 38.7 mV dec$^{-1}$ and 96 mV onset potential (at current density of 10 mA cm$^{-2}$) when the heterostructure alloy was annealed at 300 °C. These results indicate that heterostructures formed by graphene and $W_{0.4}Mo_{0.6}S_2$ alloys are far more efficient than $WS_2$ and $MoS_2$ by at least a factor of two, and they are superior than other reported TMD systems. This strategy offers a cheap and low temperature synthesis alternative able to replace Pt in the hydrogen evolution reaction (HER). Furthermore, the catalytic activity of the alloy is stable over time, i.e. the catalytic activity does not experience a significant change even after 1000 cycles. Using density functional theory calculations, we found that this enhanced hydrogen evolution in the $W_xMo_{1-x}S_2$ alloys is mainly due to the lower energy barrier created by a favorable overlap of the d-orbitals from the transition metals and the s-orbitals of $H_2$; with the lowest energy barrier occurring for the $W_{0.4}Mo_{0.6}S_2$ alloy. Thus, it is now possible to further improve the performance of the "inert" TMD basal plane via metal alloying, in addition to the previously reported strategies such as creation of point defects, vacancies and edges. The synthesis of graphene/$W_{0.4}Mo_{0.6}S_2$ produced at relatively low temperatures is scalable and could be used as an effective low cost Pt-free catalyst.

## 1. Introduction

Hydrogen, the 10[th] most abundant element on earth, has been considered as the ideal energy carrier in its molecular form ($H_2$). Unfortunately, the state-of-the-art hydrogen manufacturing, known as steam reforming from methane, results in the emission of gaseous $CO_2$[1]. In order to produce clean and renewable hydrogen efficiently, the electrolysis of water becomes an ideal route since its discovery in 1789[2,3]. The electrolysis of water involves two reactions: the hydrogen evolution reaction (HER) and the oxygen evolution reaction (OER). During the HER process, protons in the electrolyte are absorbed, and then reduced into $H_2$ on the electrode when an overpotential is applied. To reduce the overpotential, a catalyst is required to efficiently produce $H_2$. Currently, Pt is the most efficient HER catalyst due to its near-zero overpotential in acidic electrolytes[4,5]. However, the high cost and scarcity of Pt prohibits its application to fulfil the energy demand. Thus, lowering the cost of HER catalysts is of paramount importance for clean, scalable and sustainable energy.

To lower the catalysts cost, a natural abundant alternative and low-cost scalable synthesis are required. During the past years, naturally abundant $MoS_2$ coupled with other nanostructures started to gain more attention as HER catalyst,[6–8] after being ignored for their poor catalytic activity in the bulk form[9]. However, due to the semiconducting character of $MoS_2$ and $WS_2$ (2H phase), poor electrical conductivity limits the HER kinetics. In order to take advantage of the transition metal dichalcogenide (TMD) catalytic activity, conducting pathways are essential to accelerate the electron transport[7]. Phase engineering, to convert the 2H phase into a metallic 1T phase, is an alternative to increase the intrinsic electrical conductivity due to a high electron density in the d orbitals of the metal[10,11]. Besides phase engineering, the synthesis of heterostructures combining less-conducting TMDs[12] with electrical conducting materials, such as graphene, seems to show significant promise[7].

To synthesize graphene-TMD heterostructures, different synthetic approaches have been explored: chemical vapor deposition (CVD)[13–15], hydrothermal or solvothermal[7,16], and wet chemical approaches[17,18]. CVD is the most widely used method to synthesize monolayers of TMDs with high crystallinity[19–23] at high temperature (~800 °C), but the method is not ideal for the synthesis of defective catalysts that are more active, and are normally synthesized at relatively low temperature. The hydrothermal/solvothermal process is another approach able to synthesize catalytic TMDs at low temperature (~200 °C). However, the powder phase requires a binder when fabricating electrodes, thus making the processing steps less scalable, more elaborated and costly. Meanwhile, low temperature (~300 to 400 °C) wet chemical approaches based on the pyrolysis of ammonium tetrathiotungstate (ATTT, $(NH_4)_2WS_4$) or ammonium tetrathiomolybdate (ATTM, $(NH_4)_2MoS_4$), have been explored to synthesize films of $WS_2$ and $MoS_2$[17,24]. However, the aggregation of precursor molecules always gives rise to bulk phases of $XS_2$ (X=W or Mo). In order to obtain few-layers or even monolayers of $XS_2$, specific surfactants are necessary to assemble the precursor molecules into 2D domains avoiding layer stacking. In this context, graphene oxide (GO) could be used as the surfactant due to the abundant reactive functional groups, large sheet size and the ability to be transformed into conducting graphene when reduced; also known as reduced graphene oxide (rGO)[25]. Thus, monolayers or few-layers of $XS_2$ can then be assembled onto large rGO conducting sheets without aggregation after themolysis[25].

In addition to edge engineering, point defect engineering and strain engineering, some doping strategies have emerged as an alternative to obtain fast HER kinetics[26]. Transition metal doping in transition metal sulfides has been recently shown to further enhance the intrinsic catalytic activity by lowering the energy barrier of hydrogen absorption[27]. In $MoS_2$, the catalytic activity is mainly attributed to the edge sites, whereas in-plane domains are considered "inert". In order to stimulate the activity of in-plane domains, Pt atoms were doped into the $MoS_2$ to lower the in-plane energy barrier of hydrogen absorption on S atoms[28]. Therefore, we hypothesized that the in-plane doping or alloying of transition metals could be used as a rational design to activate the "inert" in-plane $MoS_2$.

In this account, we report a facile, low energy consumption, and scalable wet chemical approach able to prepare $rGO/XS_2$ (X=W, Mo or W and Mo alloy) heterostructures consisting of few layers and monolayer TMDs with conducting rGO. The resulting $rGO/XS_2$ films exhibit enhanced HER catalytic activity due to three advantages: (i) the presence of exposed edges and curved regions from dendritic-like morphologies; (ii) additional surface area from mono- and few-layered $XS_2$, and (iii) an interlayer electronic-coupling effect combining conducting rGO and catalytic $XS_2$. Most importantly, we find that $rGO/W_{0.4}Mo_{0.6}S_2$ films exhibit an outstanding HER catalyst performance with the lowest Tafel slope when compared to the un-alloyed $rGO/XS_2$ (X=W or Mo) systems, and superior to any other reported TMD system.

This HER catalytic activity was further investigated using hybrid density functional theory (DFT)[29–33]. This theoretical approach determined the chemical mechanisms for the HER by calculating the activation barrier energies, the turnover frequency (TOF), and their relationship to the electronic structure of the HER in the presence of $MoS_2$, $WS_2$, and $W_xMo_{1-x}S_2$ alloys. These models found similar Tafel slopes as those observed experimentally. In particular, we have computationally analyzed how the alloys' HER activation energy barrier is reduced when combining the Mo and W atoms in the TMD layer.

## 2. Results

The process to synthesize heterostructure $rGO/W_xMo_{1-x}S_2$ films on $Si/SiO_2$ substrates is shown in Figure 1a. Herein, GO was chosen as the template and surfactant to form edge-exposed and few-layered TMDs. First, ATTT and ATTM were dispersed into a GO aqueous solution to form a homogenous dispersion after sonication. Subsequently, the dispersion was spin-coated onto the substrate (*e.g.* $Si/SiO_2$, glassy carbon). After thermal annealing the GO is converted into rGO by restoring the π-network and by losing the oxygen functional groups via thermal reduction in the inert gas, thus $rGO/W_xMo_{1-x}S_2$ films with dendritic-like morphologies are formed, as shown by scanning electron microscopy (SEM; Figure 1b). Energy dispersive X-ray spectroscopy (EDS) mappings shown in Figures 1c-d indicate the presence of an alloy due to the homogenous distribution of W and Mo within the synthesized compound, rather than segregated phases. It is also noteworthy mentioning that the dendritic regions exhibit higher densities of W and Mo when compared to the non-dendritic sections. EDS mappings by scanning transmission electron microscopy (STEM) shown in Figure S1 also confirm the homogenous distribution of W and Mo at the nano-scale. The elemental composition of the characteristic materials obtained by EDS corresponds to: 2.85 at. % W, 3.32 at. % Mo, and 12.31 at. % S (see Figure S2), thus indicating that W to Mo ratio is 1:1.16 (x=0.46). X-ray photoelectron spectroscopy (XPS) survey spectra (Figure S3) was in good agreement with the composition obtained by EDS.

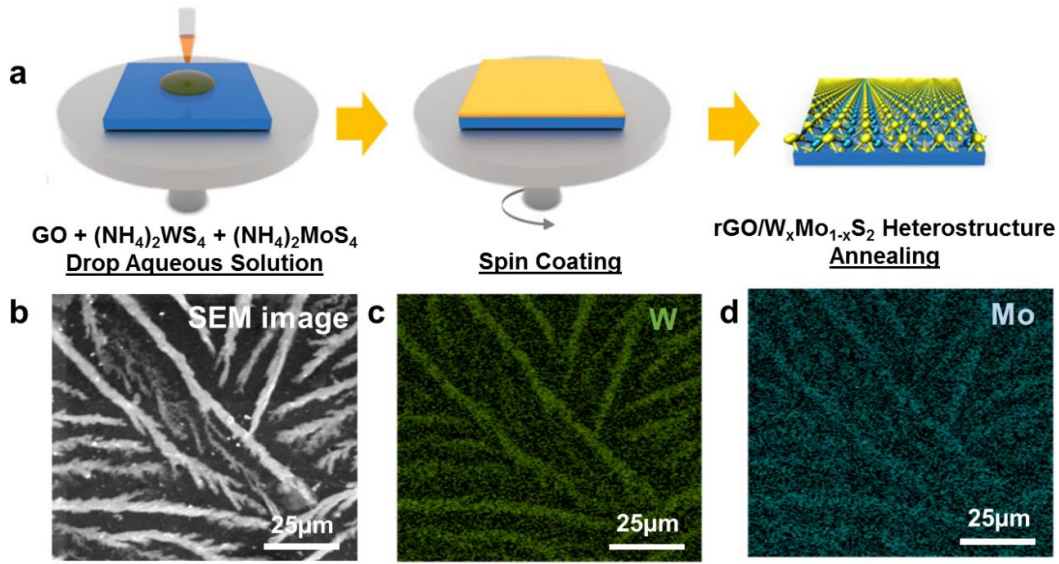

**Figure 1.** (a) Schematic representation of the wet chemical approach used to synthesize the rGO/$W_xMo_{1-x}S_2$ heterostructure starting from aqueous GO+ATTT+ATTM; (b) SEM image of the rGO/$W_xMo_{1-x}S_2$ sample on a Si/SiO$_2$ substrate; EDS mappings of: (c) tungsten (W-M line); (d) molybdenum (Mo-L line).

The structure of the rGO/$W_xMo_{1-x}S_2$ system was further studied by Raman spectroscopy (Figure 2a) using a 514.5 nm laser excitation. The synthesis and detailed characterization for rGO/WS$_2$ and rGO/MoS$_2$ can be found in the Supporting Information (Figure S4-7). For rGO/WS$_2$ films, the $A_{1g}$ and $2LA$ $(M)$ modes of WS$_2$ located at 418 cm$^{-1}$ and 352 cm$^{-1}$, respectively, can be clearly identified, as well as the $E_{2g}$ mode centered at 355 cm$^{-1}$ and overlapped with the $2LA$ $(M)$ mode. Films of rGO/MoS$_2$ (see Figure 2a) exhibit two intense bands located at 381 cm$^{-1}$and 408 cm$^{-1}$ corresponding to the $E_{2g}$ and $A_{1g}$ modes of MoS$_2$, respectively. However, the Raman spectra of the rGO/$W_xMo_{1-x}S_2$ system includes 3 significant peaks: a broad peak around 352 cm$^{-1}$ containing $2LA(M)$ and the $E_{2g}$ modes of WS$_2$; the $E_{2g}$ mode of MoS$_2$ (*ca.* 381 cm$^{-1}$); and a combined peak (*ca.* 418 cm$^{-1}$), including $A_{1g}$ modes of both $WS_2$ and $MoS_2$. Raman mappings of the *2LA*, *$E_{2g}$* and *$A_{1g}$* modes for $WS_2$ and $MoS_2$ (see Figure S8) further illustrate the homogenous distribution of W and Mo atoms within the produced films, and provide more evidence to support the formation of $WS_2$ and $MoS_2$ alloys. The *2LA* ($WS_2$)/*$E_{2g}$* ($MoS_2$) intensity ratio mapping was also captured to further illustrate the distribution of W and Mo atoms (Figure S8f). Most of the analyzed area exhibits equivalent distribution of W and Mo atoms.

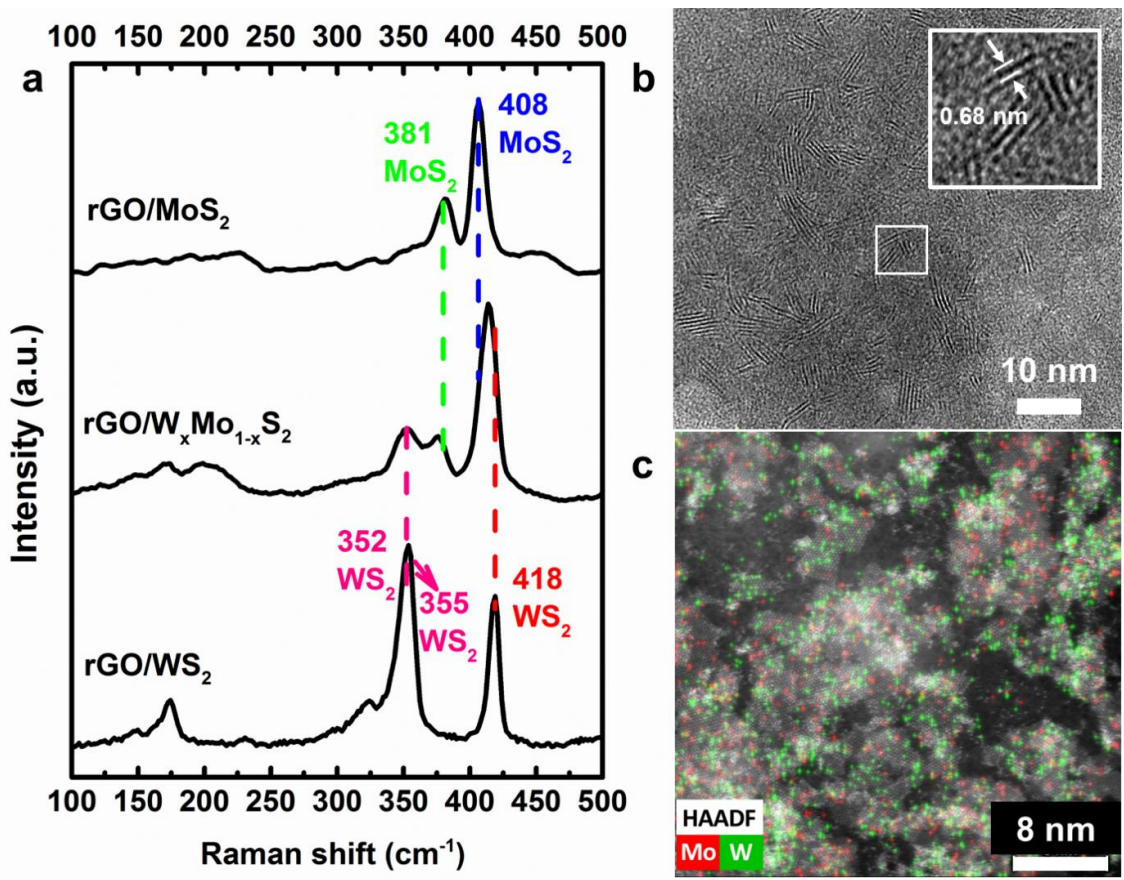

**Figure 2.** (a) Raman spectra of rGO/$WS_2$, rGO/$W_xMo_{1-x}S_2$, and rGO/$MoS_2$; (b) High-resolution transmission electron microscopy (HRTEM) image of the rGO/$W_xMo_{1-x}S_2$ film within the dendritic area; (c) High magnification STEM- high angle annular dark field (HAADF) image with overlapped EDS mappings showing areas of rGO/$W_xMo_{1-x}S_2$ at the non-dendritic area.

A representative high-resolution transmission electron microscopy (HRTEM) image (Figure 2b) of the dendritic area demonstrates dense nanostructures constructed from vertically aligned few-layered $W_xMo_{1-x}S_2$, exhibiting an interlayer spacing of 0.68 nm, which is slightly higher than the d (002) of 2H-WS$_2$ and 2H-MoS$_2$ (~0.62 nm). This enlarged interlayer spacing could be due to the distortion and structural defects present within the alloys[34]. The $W_xMo_{1-x}S_2$ nanostructures with small crystal domains and few-layers are attributed to the self-assembly of oxygen functional groups from GO, as GO offers a large curved surface template able to arrange anions of $WS_4^{2-}$ and $MoS_4^{2-}$, and avoids agglomeration while undergoing thermolysis[25]. Beyond the dendritic-area, small size monolayer $W_xMo_{1-x}S_2$ fragments were observed via STEM with the high angle annular dark field (HAADF) detector in the non-dendritic regions (Figure 2c). These fragments are partially overlapped and stacked, thus forming few-layer structures. The EDS mappings provides further evidence that W and Mo are well mixed within the basal planes (Figure 2c).

## 3. Discussion

After thermal annealing, the GO is converted to conducting rGO by removing oxygen functional groups and restoring the $\pi$-conjugation of rGO. Due to the formation of mono- and few-layer TMDs on rGO, abundant edges from dendritic-like regions are produced, thus making these TMD/rGO films attractive for catalysis if deposited on conducting substrates. Therefore, the HER activity was studied and compared between rGO/MoS$_2$, rGO/WS$_2$, and rGO/$W_xMo_{1-x}S_2$ films on glassy carbon in 0.5 M $H_2SO_4$ electrolyte (experimental details about the HER measurements are described in the methods section). As shown in Figure 3a, the glassy carbon substrate and rGO exhibits negligible HER activity according to the polarization curves. However, once the rGO/TMDs heterostructure films are coated on the glassy carbon substrates, significant HER activities are noted: 302 mV, 279 mV, and 233 mV for rGO/MoS$_2$, rGO/WS$_2$, and rGO/W$_x$Mo$_{1-x}$S$_2$, respectively, when the current density is 10 mA cm$^{-2}$. In order to evaluate the inherent reactivity of the catalysts, the Tafel plots are linearly fitted (see Figure 3b), yielding 111.4 mV dec$^{-1}$ and 81.3 mV dec$^{-1}$ for rGO/MoS$_2$ and rGO/WS$_2$; these values are significantly higher than the rGO/W$_x$Mo$_{1-x}$S$_2$ alloy (50.6 mV dec$^{-1}$) when annealed at 400 °C.

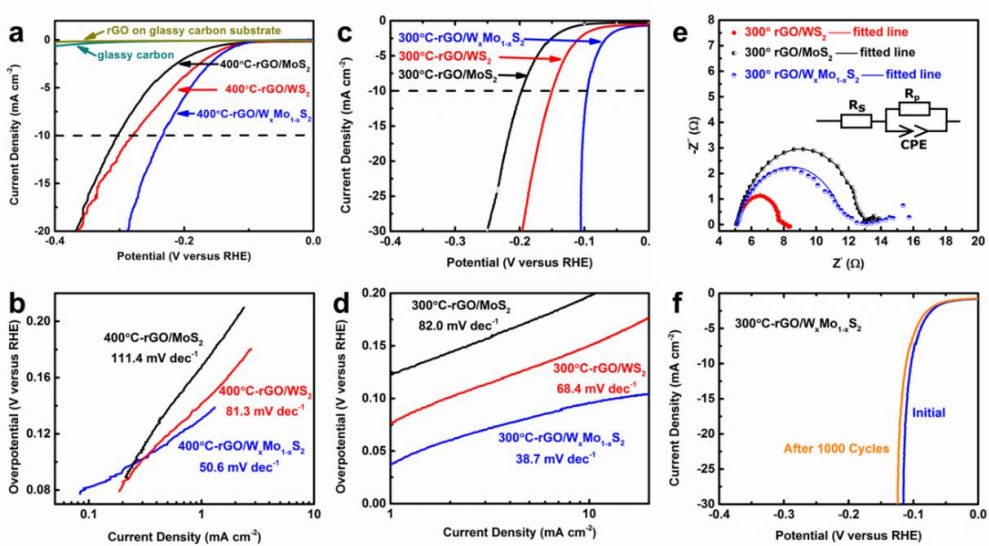

**Figure 3.** (a) Polarization curves (I-V) of glassy carbon, rGO, rGO/MoS$_2$, rGO/WS$_2$, rGO/W$_x$Mo$_{1-x}$S$_2$ films (annealed at 400 °C) on glassy carbon, and (b) corresponding Tafel plots; (c) Polarization curves (I-V) of rGO/MoS$_2$, rGO/WS$_2$, rGO/W$_x$Mo$_{1-x}$S$_2$ films (annealed at 300 °C) on glassy carbon, and (d) corresponding Tafel plots; (e) Electrochemical impedance spectroscopy(EIS) Nyquist spectra of rGO/W$_x$Mo$_{1-x}$S$_2$ film on glassy carbon (annealed at 300 °C) and the fitting circuit is in the inset; (f) polarization curves (I-V) of rGO/W$_x$Mo$_{1-x}$S$_2$ film on glassy carbon (annealed at 300 °C) at the first cycle and after 1000 cycles.

We also found an improved HER activity in the rGO/$W_xMo_{1-x}S_2$ alloy annealed at 300 °C as seen in Figure 3c. In particular, rGO/$W_xMo_{1-x}S_2$ films annealed at 300 °C exhibit the lowest onset potential of 96 mV (at 10 mA cm$^{-2}$), when compared to 197 mV for rGO/$MoS_2$ and 150 mV for rGO/$WS_2$. The Tafel slope for rGO/$W_xMo_{1-x}S_2$ films annealed at 300 °C is 38.7 mV dec$^{-1}$, which is still higher than Pt (~30 mV dec$^{-1}$)[35] because of the different reaction mechanism[35]. Pt normally follows the Volmer reaction mechanism while metal sulfides mainly follows the Heyrovsky reaction mechanism. However, it is far superior than other rGO/$MoS_2$ and rGO/$WS_2$ films (annealed at 300 °C or 400 °C), and any other reported TMD system listed in Table 1. In this context, it has been discovered that by decreasing the annealing temperature from 400 °C to 300 °C, extra S ligands and small domain sizes were obtained with Mo/S stoichiometries close to 2.19[36], which is also confirmed by the XPS analysis. Deconvoluted Mo 3d spectra (Figure S9a, d) exhibits multiple peaks. The peaks at 229.0 and 232.0 eV correspond to $Mo^{4+}$ $3d_{5/2}$ and $3d_{3/2}$ orbitals of 2H-$MoS_2$[37,38]. Additional peaks are observed at 232.6 eV ($Mo^{6+}$ $3d_{5/2}$) and 235.6 eV ($Mo^{6+}$ $3d_{3/2}$) assigned to $MoO_3$; 229.6 eV ($Mo^{5+}$ $3d_{5/2}$) and 233.3 eV ($Mo^{5+}$ $3d_{3/2}$) assigned to $Mo_2S_5$[36]. It is clear that the increase of $Mo_2S_5$ was found in rGO/$MoS_2$ annealed at 300 °C. Moreover, the line width of $MoS_2$ peaks becomes broader in rGO/$MoS_2$ annealed at 300 °C. Similarly, in the S 2p spectrum, doublet peaks of 2H-MoS2 appear at 162.9 eV ($S^{2-}$ $2p_{3/2}$) and 161.8 eV ($S^{2-}$ $2p_{1/2}$; Figure S9 b, e)[37]. Besides Mo 3d and S 2p peaks, the C 1s peaks shown in Figure S9 c and f demonstrate the highest intensity at 284.4 eV, corresponding to C=C/C-C in aromatic rings[39]. A side peak located at 285.1 eV indicates the C-N bond because of the presence of N in ATTT/ATTM. The fraction of C-O peak can be barely observed, indicating the formation of rGO after 300 °C or 400 °C thermal annealing.

The Raman signal of the films annealed at 300 °C (Figure S10) showed enhanced peaks close to the *LA(M)* when compared to films annealed at 400 °C (Figure 2a). Since the increase in intensity close to the *LA(M)* peak is correlated to the increase in structural disorder, similar as the D-band for graphene[40], the overall improved catalytic performance of the films synthesized at 300 °C is due to a higher degree of structural disorder and the presence of extra sulfur atoms[41].

**Table 1.** Summary of state-of-art $WS_2$ and $MoS_2$ related materials performance for HER is shown, and they are compared to the materials reported in this work.

| Samples | Method | Temp. (°C) [a] | Onset potential (mV) [b] | Tafel slope (mV dec$^{-1}$) |
|---|---|---|---|---|
| Defect-rich $MoS_2$[42] | Hydrothermal | 220 | -200 | 50 |
| 1T $MoS_2$[10] | Li intercalation | 25 | -200 | 40 |
| $MoS_x$ on graphene protected Ni foam[43] | CVD; wet chemical route | 1050; 100 to 300 | -240 | 42.8 |
| Vertically aligned $MoS_2$ and $MoSe_2$[6] | CVD | 550 | <-400 | 105-120 |
| $MoS_2$ on Au[44] | PVD/CVD | 400 to 550 | <-200 | 55-60 |
| 1T $WS_2$[11] | Li intercalation | 25 | ~-200 | 55 |
| $MoS_2$ quantum dots[45] | Solvent exfoliation | 140 | ~-250 | 115 |
| $WS_2$ quantum dots[45] | | | ~-350 | 138 |
| rGO/$MoS_2$[7] | Hydrothermal | 200 | ~-150 | 41 |
| rGO/$WS_2$[16] | Hydrothermal | 265 | ~-250 | 58 |
| MoSSe[46] | Wet chemical route | 400 | ~-164 | 48 |
| rGO/$W_xMo_{1-x}S_2$ (This work) | Wet chemical route | 300 | -96 | 38.7 |
| rGO/$MoS_2$ (This work) | | | -197 | 82.0 |
| rGO/$WS_2$ (This work) | | | -150 | 68.4 |
| rGO/$W_xMo_{1-x}S_2$ | | | -233 | 50.6 |
| (This work) | | | | |
| rGO/MoS$_2$ (This work) | | 400 | -302 | 111.4 |
| rGO/WS$_2$ (This work) | | | -279 | 81.3 |

[a] The temperature here is referred to the temperature at which the synthesis was carried out.
[b] The onset potential reported is at 10 mA cm$^{-2}$.

In order to get further insight into the kinetics of the HER process, we performed electrochemical impedance spectroscopy (EIS) when the overpotential is 0.1 V. To model the electrochemical reaction near the electrode surface, the EIS plots were fitted using a Randles circuit in which $R_s$ is the uncompensated solution resistance, $R_p$ is the polarization resistance representing the charge transfer resistance of the electrode, and CPE is a constant phase element used to calculate the double layer capacitance; the fitted results are listed in Table S1. Based on Figure 3e and Table S1, the charge transfer resistance ($R_p$) of all three films are of the same magnitude and ranged from 2.69 to 7.79 $\Omega$. The small charge transfer resistance can be attributed to the stacked structure of TMDs and rGO, and exposed edges from vertically aligned heterostructures, thus ensuring the isotropic electron transport between glassy carbon and the films[34]. The large $C_{dl}$ of the films (listed in Table S1) indicate high exposure of the active surface, since $C_{dl}$ is proportional to the effective electrochemically active surface area[47]. It is clear that due to the fast charge transfer and highly exposed surface area compared to other studies[34], improved HER activities were obtained for rGO/MoS$_2$, rGO/WS$_2$, and rGO/ W$_x$Mo$_{1-x}$S$_2$ films. However, the $R_p$ and $C_{dl}$ values of the most efficient catalyst, rGO/W$_x$Mo$_{1-x}$S$_2$, are not the most prominent among the three films; they are within the same magnitude. Thus, charge transfer and surface area are not the main factors responsible for improving the catalytic performance of rGO/$W_xMo_{1-x}S_2$ alloys, even after 1000 cycles (Figure 3f), but these are some of the reasons for the overall improved catalytic performance of rGO/$XS_2$ (X=W, Mo, or W and Mo alloy) systems synthesized by this approach. In order to further elucidate the mechanism of the improved HER activity in rGO/$W_xMo_{1-x}S_2$ alloys, STEM and DFT calculation were performed.

Thus, STEM observation was carried out to visualize the heterostructure of rGO/$W_xMo_{1-x}S_2$ to elucidate the improved catalytic activity of rGO/$W_xMo_{1-x}S_2$. In particular, a HAADF detector was used for Z-contrast imaging. Since the atomic Z-number for W (74) and Mo (42) are different, the distribution of W and Mo atoms in the $XS_2$ structure was clearly visualized using this technique. Various other local structures were also observed by STEM. For example, local alloying composition changes of the metal atoms were observed, i.e. Mo present within the $WS_2$-rich area (Figure 4a), and W present in the $MoS_2$-rich area (Figure 4b). Interestingly, $WS_2$ and $MoS_2$ triangular domains were also found in Figure 4c and Figure 4d, respectively. The lattice constants (*a* and *b*) of pristine $WS_2$ and $MoS_2$ are slightly different, thus the alloy formation can induce local strain within the 2D lattice. In order to reduce strain inside the film, metal atoms could be segregated and then allowed to form triangular domains during annealing. Other frequently observed structures were W-decorated $MoS_2$ (Figure 4e), and extended triangular vacancies (Figure 4f). Thus, the coexistence of W and Mo introduces the presence of more exposed edges, which are responsible for altering the atomic environment of the metal and influencing the hydrogen absorption within the basal plane.

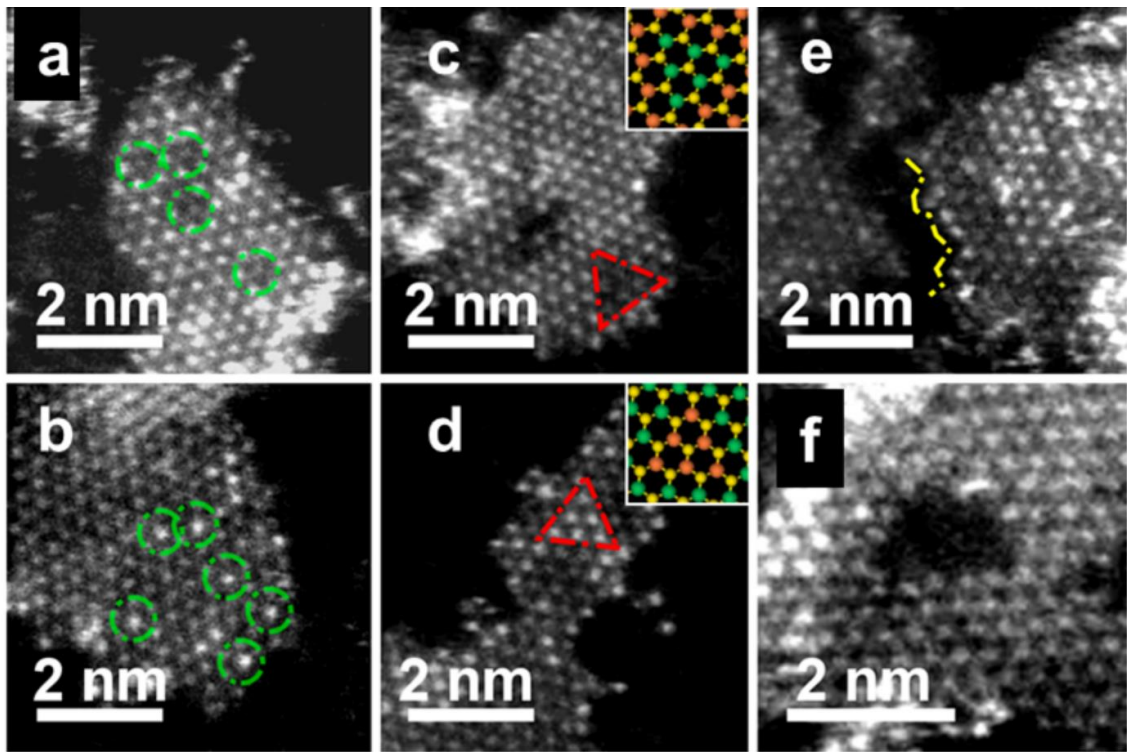

**Figure 4.** High magnification STEM-HAADF image of the rGO/$W_xMo_{1-x}S_2$ heterostructure. The intensity of image correlated to Z-number of element thus the brighter and darker dots correspond to tungsten (W) and molybdenum (Mo) atoms, respectively. The metal atom substitution was observed in (a) $WS_2$-rich and (b) $MoS_2$-rich areas. Both of (c) Mo and (d) W triangular domains were also observed, as well as (e) W-decorated $MoS_2$ and (f) vacancy.

In addition, we have investigated the catalytic activity of the $W_xMo_{1-x}S_2$ alloys in the presence of rGO by computing the electronic properties of the layered system depicted in Figure 5a. The results of the $MoS_2$ with rGO and $WS_2$ with rGO are shown in Figure S11. These calculations show that both the pristine monolayer $MoS_2$ and $WS_2$ are pure semiconductors with band gap of ~2.5 eV. Due to addition of rGO, the band gaps of both the pristine and alloys were decreased to about 1.30 eV, as depicted in the band structures and density of states (DOSs) calculations shown in Figure 5a and Figure S11. Thus, the rGO affects the electronic properties of the pristine and alloyed heterostructures studied here. These calculations demonstrate that there is an interlayer electronic-coupling between rGO and both the pristine $MoS_2/WS_2$ and $W_xMo_{1-x}S_2$ alloys that decreases the band gap. The addition of rGO to the TMDs changes the electron accumulation in the conduction and valence bands as shown in the DOSs suggesting high electron mobility. However, these findings do not explain the superior performance of the alloy over pristine TMDs, but rGO helped to improve the conductivity of TMDs as we initially hypothesized. In an analog heterostructure, Zheng *et al.*[48] reported that the rGO/$C_3N_4$ heterostructure exhibited a higher activity for HER than the pristine $C_3N_4$. To explain the effect of rGO on the catalytic system, Zheng *et al.* studied only the electronic ground state properties (*i. e.* the DOS and the band structure) and infer that the rGO is helping the reaction somehow, however the full reaction mechanism was not studied. In contrast, we believe that the study of reaction barriers is needed to explain the chemical reactivity of the catalyst, instead of extrapolating the kinetic effects from the electronic ground state properties which explains only the equilibrium thermodynamics.

Therefore, to further describe the HER mechanisms and the catalytic activity of $W_xMo_{1-x}S_2$ alloys, we have considered different reactions pathways with the most prominent being two reaction steps: (i) H atom migrates from the S atom to the transition metal (W or Mo) atom, i.e. H· migration, which is known as the Volmer reaction mechanism, and (ii) $H_2$ formation where one absorbed hydride (H⁻) reacts with a solvated proton of an adjacent explicit water, also known as the Heyrovsky reaction mechanism. Thus, in the Volmer reaction mechanism, the rate determining step is the migration of a hydrogen atom, whereas in the Heyrovsky reaction mechanism, the rate determining step requires an adjacent hydronium ($H_3O^+$), which is the source of a proton along the reaction pathway involving adsorbed H· atoms to form $H_2$ (see Figure S12). It was found that these two reaction pathways have the largest reaction barriers for HER of the pristine and alloyed materials. The energy barriers are shown in Figure 5b-c, Table S2 and Table S3. The calculations also revealed that the catalytic activity of the alloy ($W_xMo_{1-x}S_2$) is higher than that of pristine $WS_2$ or $MoS_2$ phases, which agrees well with the experimental observations (Figure 3a-d). In addition, the experimental Tafel slope is lower for the $W_xMo_{1-x}S_2$ alloy (Figure 3b, d), and the calculations indicate that this is true for the alloy system regardless of the level of transition metal (TM) substitution due to the presence of a better overlap for the d-orbitals from the transition metal and the s-orbitals of $H_2$. As a result, the basal planes of the TMDs, which were considered more "inert" compared to the edges[44], get activated via the formation of TMDs alloys.

More interestingly, DFT[29-32] calculations revealed that the $W_{0.4}Mo_{0.6}S_2$ alloy has the lowest reaction energy barriers for both the H· migration and $H_2$ formation, even when compared to pristine $WS_2$ and $MoS_2$ materials (Figure 5b-c). More specifically, the present computation indicates that the activation energy barrier in the solvent phase for the H· migration reaction in $W_{0.4}M_{0.6}S_2$ alloys is 11.9 kcal $mol^{-1}$, while for the pristine $MoS_2$ and $WS_2$ materials the activation energy barriers are 17.7 kcal $mol^{-1}$ and 18.1 kcal $mol^{-1}$, respectively (see Figure 5b and Table S2). Additionally, the activation energy barrier for $H_2$ evolution following the Heyrovsky reaction mechanism of the $W_{0.4}M_{0.6}S_2$ alloy is 13.3 kcal $mol^{-1}$, compared to 23.8 kcal $mol^{-1}$ and 21.3 kcal $mol^{-1}$ for the pristine $MoS_2$ and $WS_2$ materials, respectively (see Figure 5c, and Table S3). The present computation reveals that this Heyrovsky activation energy barrier for $H_2$ evolution reaction of the $W_{0.4}M_{0.6}S_2$ alloy is about ~ 7.6 kcal mol$^{-1}$ smaller than Pt(111) catalyst reported by Skulason *et al.*[49] and Fang *et al.*[50] Some insight of the electronic role in the mechanism can be understood from the highest occupied molecular orbital (HOMO) calculations of the transition state (TS) structures depicted in Figure 5d-f. In this context, the transition state of the rate limiting reaction step (the Heyrovsky mechanism) is stabilized by the better overlap of the d-orbitals of the transition metal and the s-orbitals of the $H_2$ molecules. This stabilization of the reaction limiting step is crucial for optimizing the reaction barriers, thus the overall catalysis. This mechanism provides key insight on why $W_xMo_{1-x}S_2$ alloys is an effective catalyst for HER, i.e. the tuning of the d-orbital of the TMDs overlaps with the s-orbitals $H_2$ is of the most importance. This new strategy is different from the well-known approaches used for tuning the $H_2$ binding energy to TMDs or the control of the acidity of the proton source.

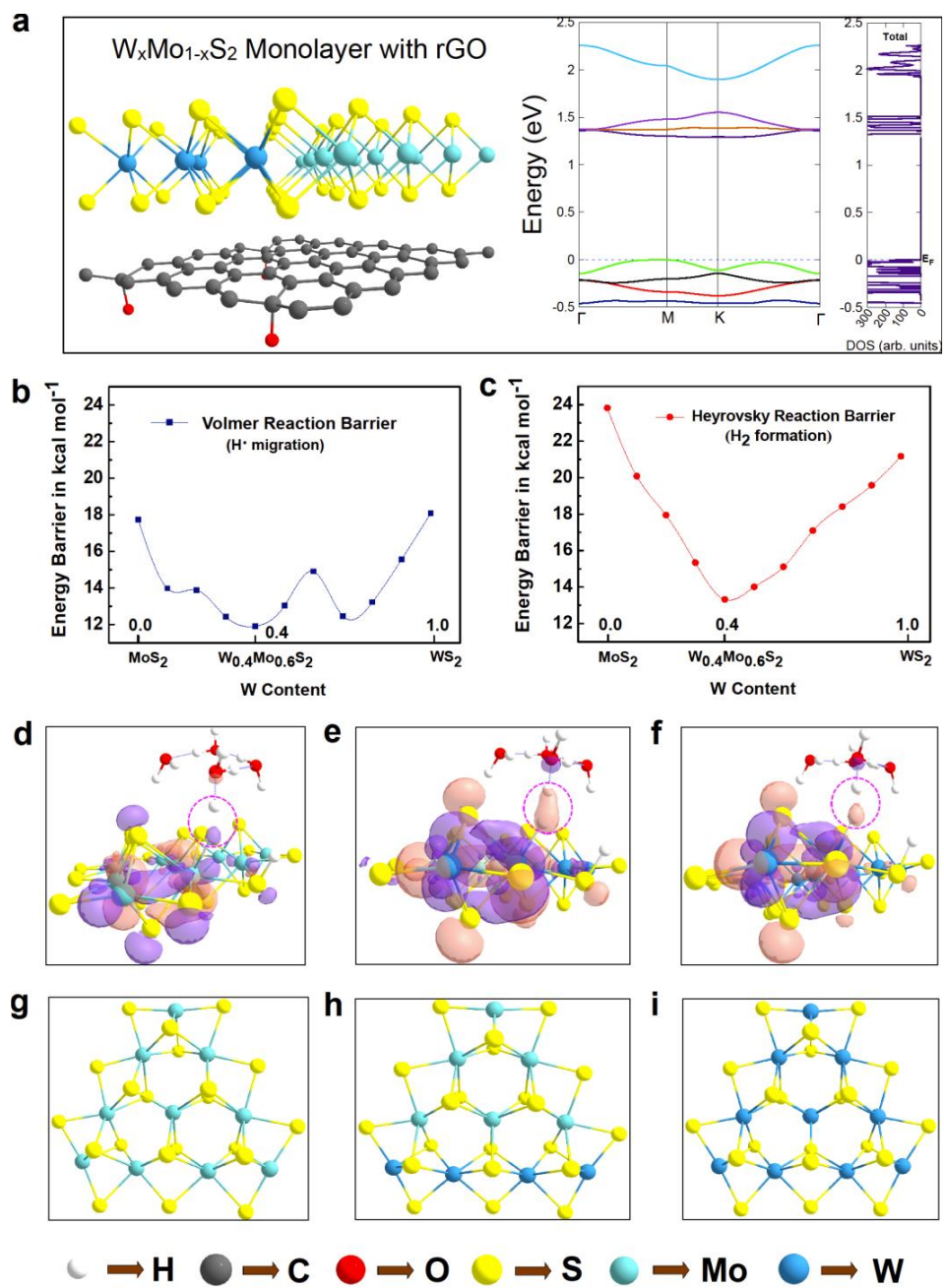

**Figure 5.** The effect of alloying on (a) the band structure and DOSs for the rGO/$W_{0.4}Mo_{0.6}S_2$ heterostructure, (b) the activation energy barrier for the migration of hydrogen atom (Volmer reaction mechanism); and (c) the activation barrier for the formation of $H_2$ molecule involving adjacent water (Heyrovsky reaction mechanism). The HOMO of the transition state structures in the Heyrovsky reaction path are shown for (d) $MoS_2$, (e) $W_{0.4}Mo_{0.6}S_2$ and (f) $WS_2$, where the molecular orbitals involved in the $H_2$ formation are highlighted by a pink circle. The optimized geometries of catalysts are shown for the (g) $MoS_2$, (h) $W_{0.4}Mo_{0.6}S_2$, and (i) $WS_2$.

To further compare with our experimental results, we have theoretically computed the Tafel slope $b$ of the proposed HER mechanism in the $W_{0.4}Mo_{0.6}S_2$ system, assuming the electron transfer from the support to the catalyst does not limit the rate. Interestingly, we found that the theoretical value of the Tafel slope $b$ for this reaction is 59.0 mV dec$^{-1}$, which agrees with the experimental value of 50.6 mV dec$^{-1}$ when the annealing temperature is 400 °C. The experimental Tafel slope is 38.7 mV dec$^{-1}$ when films were annealed at 300 °C. This suggests that, besides the alloy formation, extra S atoms (Figure S3) and structural disorder result in the decrease in Tafel slope of the sample annealed at 300 °C. Table S3 indicates that TOF is higher for all the W/Mo alloys than the pristine compounds, especially when the W to Mo ratio is between 1:1 (x=0.48) to 1:1.5 (x=0.40). The calculated TOF of the $W_{0.4}Mo_{0.6}S_2$ alloy is about $1.1 \times 10^3$ sec$^{-1}$, which is the highest among all the alloy $W_xMo_{1-x}S_2$ systems, including the pristine $MoS_2$ and $WS_2$ (Table S3). Meanwhile, our experimental EDS results shown in Figure S2 indicate that the W to Mo ratio is 1:1.16 (x=0.46), within the range of 1:1 (x=0.48) to 1:1.5 (x=0.40), thus confirming the theoretical prediction and further validating the proposed chemical reaction mechanism.

## 4. Conclusion

In summary, graphene/$W_xMo_{1-x}S_2$ heterostructure films with dendritic-like morphology were synthesized by a facile and scalable wet chemical approach followed by annealing at 300-400

°C. It was found that the formation of W-Mo alloys ($W_{0.4}Mo_{0.6}S_2$) can significantly enhance the HER ability compared to pure $WS_2$ and $MoS_2$ phases. The alloys annealed at 300 °C are reported to have a 38.7 mV dec$^{-1}$ Tafel slope and 96 mV onset potential (at current density 10 mA cm$^{-1}$), while the alloys annealed at 400 °C has a Tafel slope 50.6 mV dec$^{-1}$. DFT calculations indicate that the high reactivity observed in these TMD alloys ($W_xMo_{1-x}S_2$), when compared to pure phases ($MoS_2$ or $WS_2$), is due to the lowering of the activation energy barrier when forming $H_2$ molecules along the "inert" basal planes. This low activation barrier is due to the stabilization of the rate determinant transition state, where the electron density of $H_2$ formation is favored by the overlap of the s-orbitals of the H atoms and the d-orbitals from the transition metals alloys from W and Mo. In other words, this electronic overlap stabilizes the transition state, which in consequence lowers the Tafel slope, thus making the alloys responsible for a better catalytic activity regardless of the alloying concentration. The lowest energy barrier for $W_xMo_{1-x}S_2$ can be found when x=0.4, which is in good agreement between the theoretical and experimental data. In conclusion, the catalytic activity for $H_2$ evolution can be tuned by substituting transition metals forming alloys, thus the "inert" basal planes can be activated. It is therefore clear that these unique layered heterostructures possess enormous potential in catalysis, and that the current work will trigger further complementary studies.

## 5. Experimental Section

*Synthesis of graphene oxide*

In order to obtain few layered and large GO flakes, graphite flakes (purchased from Asbury Co.) were processed by a modified Hummers method[51]. Briefly, graphite (2.5 g) and potassium nitrate (KNO$_3$, 2.0 g) were mixed with sulfuric acid (H$_2$SO$_4$, 130 ml) in an ice bath. Then potassium permanganate (KMnO$_4$, 12.0 g) was added slowly into the slurry. After stirring for 2 h under an ice bath, the mixture was kept stirring at room temperature for 5 days. Then the slurry was heated to 98 °C and slowly added to boiling water (300 ml). After further stirring for 2 h at 98 °C, the mixture was cooled down to 60 °C and hydrogen peroxide (H$_2$O$_2$, 35 % in aqueous solution, 8 ml) was added into the mixture drop by drop. Subsequent stirring for another 2 h at room temperature was carried out and the resulting material was washed with hydrochloric acid (HCl, 10 % in aqueous solution) three times, and then washed with deionized water by centrifugation until it reached a pH value of 5. Finally, the GO solution was diluted to 3.5 mg ml$^{-1}$ for further synthesis.

*rGO/XS$_2$ synthesis*

Silicon wafers with a 285 nm SiO$_2$ layer were cleaned with piranha solution (3:1 H$_2$SO$_4$:H$_2$O$_2$ mixture) for 30 min, and then cleaned with acetone and 2-propanol in an ultrasonic bath for 20 min. For the synthesis of rGO/XS$_2$ (X=W or Mo), ATTT or ATTM (7 mg) were mixed with GO aqueous solution (2 ml, 3.5 mg ml$^{-1}$). For the alloy sample, ATTT (3.5 mg) and ATTM (3.5 mg) were mixed with GO solution (2 ml, 3.5 mg ml$^{-1}$). After 10 min of ultra-sonication, the mixture was spin-coated on the pre-cleaned Si/SiO$_2$ substrate for 30 s at 1000 rpm, as shown in Figure 1a. Subsequently, the coated substrate was heated on the hot plate at 120 °C to evaporate the excess water. Then, the samples were loaded into a quartz tube for annealing at 120 °C for 15 min for removing the water molecules among the GO layers, and then annealed at 400 °C (or 300 °C) for 30 min under Ar/H$_2$ (500 sccm, 15 % H$_2$) flow. For the HER catalytic studies, films on glassy carbon substrates were prepared, with the area of the film being 1 cm by 1 cm.

*Electrochemical measurement*

HER measurements were performed in a 3-electrode system with a graphite rod as counter electrode, and Ag/AgCl (3 M NaCl) as reference electrode using a Versa STAT 4 potentiostat. The reference electrode was calibrated with respect to the reversible hydrogen electrode (RHE) with high purity $H_2$ saturated $H_2SO_4$ (0.5 M) electrolyte. The scan rate of linear sweep voltammetry was 1 mV s$^{-1}$. EIS was performed in the same configuration at overpotential of 100 mV, and the frequency range is 1 MHz to 10 mHz at the amplitude of sinusoidal voltage of 10 mV. Software Zview was used to fit the EIS Nyquist plots, and $C_{dl}$ was calculated by the equation

$$C = R^{\frac{1-n}{n}} Q^{\frac{1}{n}}$$

where R is $R_p$, n=CPE-P, and Q=CPE-T[52].

*Computational Details and Theoretical Calculations*

We performed a hybrid dispersion-corrected DFT-D (here B3LYP-D2)[53–57] calculations using CRYSTAL14[58] to investigate the effects of rGO on both the pristine and alloy layer structures (Figure 5a and Figure S11). Semi-empirical Grimme's (-D2) dispersion corrections were added in the present calculations in order to incorporate van der Waals (vdW) dispersion interactions in the system[55-57, 59-60]. Triple-zeta valence with polarization function quality (TZVP) basis sets were used for the C, O and S atoms[61], and HAYWSC-311(d31)G type basis set with Hay-Wadt type effective core potentials (ECPs) for both the Mo[62] and W[63]. All integrations of the first Brillouin zone were sampled on a 4x4x1 Monkhorst-Pack[64] k-mesh grids.

A cluster model system was used to investigate computationally the HER mechanism including the reaction barriers as shown in Figure 5d-i. The cluster model system contains the general formula $W_xMo_{1-x}S_2$, where x=1 and 0 for the pristine compounds and x=0.1-0.9 for the alloys. The $W_xMo_{1-x}S_2$ alloys have been prepared by substituting Mo atoms by W atoms in $MoS_2$ in different configurations. To study the HER mechanism, the DFT-M06L[29-30] method was used. It has been reported that the DFT M06L gives reliable energy barriers for reaction mechanisms of organometallic catalysts[30]. For this model, we have used the 6-31+G** basis sets for H, S, and O atoms[65,66], while LANL2DZ basis sets with ECPs for Mo and W atoms[67]. We used the polarizable continuum model (PCM) for all the calculations (including optimization of reactants and transition states) to capture solvation effects in the DFT calculations, while four water molecules were added explicitly for the Heyrovsky reaction mechanism. For the PCM calculation, a dielectric constant of 80.13 for water was used. The optimized geometries of these materials ($MoS_2$, $WS_2$, and $W_xMo_{1-x}S_2$) and transition structures are shown in Figure S13-14. The HOMO and the lowest unoccupied molecular orbital (LUMO) of the TS1, TS2 and TS3 involved in the reactions are shown in Figure S15-16. The vibrational frequencies were computed at the optimized geometry to obtain the zero-point vibrational energy (ZPE) at the same level of theory. The transition states were confirmed by observing only one imaginary frequency in the vibrational modes. All the computations for the HER mechanism were performed with the general-purpose electronic structure quantum chemistry program *Gaussian09*[68] to obtain the optimized geometries and transition structures for this model.

## Supporting Information

Characterizations (STEM images, EDS spectrum, SEM images, Raman spectra and mapping, electrochemical impedance spectroscopy) for rGO/$WS_2$ and rGO/$MoS_2$, as well as electronic structure (LUMO and HOMO), optimized geometries of the reaction mechanism, and reaction barriers are supplied as Supporting Information. The Supporting Information is available free of charge on the ACS Publication website.

## Acknowledgments

This work is supported by the U.S. Army Research Office MURI grant W911NF-11-1-0362. We also thank the National Science Foundation: 2DARE-EFRI 1542707 (MT) and EFRI-1433311 (ZL, CJ, ALE). Y.L. acknowledges the support of China Scholarship Council. R.L. acknowledges the support from the National Natural Science Foundation of China (Grant No. 51372131, 51232005), 973 program of China (No. 2014CB932401, 2015CB932500), and Beijing Nova Program and the Tsinghua University Initiative Scientific Research Program. The authors are also grateful for the support of the material characterization laboratory (MCL) at Pennsylvania State University for SEM, EDS, and TEM characterizations. The authors acknowledge Jeff Shallenberger for the XPS measurements. S.P., O.O.I. and J.L.M-C. were supported by Florida State University (FSU). J.L.M-C. and S. P. acknowledge the support from the Energy and Materials Initiative at FSU. The computing for this project was performed on the High Performance Computer cluster at the Research Computing Center at FSU. S.P. acknowledges Kevin P. Lucht and Yohanes Pramudya, FSU for helpful discussions and guidance with computational resources. The authors would like to thank the anonymous reviewers for their valuable comments and suggestions to improve the quality of the manuscript.

# Supporting Information

# Low Temperature Synthesis of Heterostructures of Transition Metal Dichalcogenide Alloys (W$_x$Mo$_{1-x}$S$_2$) and Graphene with Superior Catalytic Performance for Hydrogen Evolution

Yu Lei[1†], Srimanta Pakhira[2,3,4,5†], Kazunori Fujisawa[6], Xuyang Wang[7], Oluwagbenga Oare Iyiola[2,3,4,5], Néstor Perea López[6,8], Ana Laura Elías[6,8], Lakshmy Pulickal Rajukumar[1,8], Chanjing Zhou[1], Bernd Kabius[1], Nasim Alem[1], Morinobu Endo[9], Ruitao Lv[7*], Jose L. Mendoza-Cortes[2,3,4,5*], and Mauricio Terrones[1,6,8,9,10*]

[1]Department of Materials Science and Engineering & Materials Research Institute, The Pennsylvania State University, University Park, Pennsylvania 16802, United States; [2]Department of Chemical & Biomedical Engineering, FAMU-FSU College of Engineering, Florida State University (FSU), Tallahassee, Florida, 32310, United States; [3]Department of Scientific Computing, 400 Dirac Science Library, FSU, Tallahassee, Florida, 32304, United States; [4]Materials Science and Engineering Program, High Performance Materials Institute; FSU, Tallahassee, Florida, 32310; [5]Condensed Matter Theory, National High Magnetic Field Laboratory (NHMFL), FSU, Tallahassee, Florida, 32310, United States; [6]Department of Physics, The Pennsylvania State University, University Park, Pennsylvania 16802, United

States;  [7]Key Laboratory of Advanced Materials (MOE), School of Materials Science and Engineering, Tsinghua University, Beijing, 100084, China; [8]Center for 2-Dimensional and Layered Materials, The Pennsylvania State University, University Park, Pennsylvania 16802, United States; [9]Institute of Carbon Science and Technology, Shinshu University, Wakasato 4-17-1, Nagano, 380-8553, Japan; and [10]Department of Chemistry, The Pennsylvania State University, University Park, Pennsylvania 16802, United States. [†]The authors Y.L. and S.P. contributed equally in this article. Correspondence and requests for materials should be addressed to R.L, J.L.M-C. or M.T. (email: lvruitao@tsinghua.edu.cn, jmendozacortes@fsu.edu, or mut11@psu.edu).

**Characterization**

The scanning electron microscopy (SEM) was carried out using a FEI Nova NanoSEM 630 FESEM equipped with an energy dispersive spectroscopy (EDS).

Raman spectroscopy and mapping were performed in a Renishaw inVia confocal microscope-based Raman spectrometer with a spectral resolution better than 2 $cm^{-1}$. We used a 514.5 nm laser excitation, and the 520.5 $cm^{-1}$ phonon mode from the silicon substrate was used for calibration.

X-ray photoelectron spectroscopy (XPS) experiments were performed using a Physical Electronics VersaProbe II instrument equipped with a monochromatic Al kα x-ray source (hv = 1486.7 eV) and a concentric hemispherical analyzer. Charge neutralization was performed using both low energy electrons (<5 eV) and argon ions. The binding energy axis was calibrated using sputter cleaned Cu foil (Cu $2p_{3/2}$ = 932.7 eV, Cu $2p_{3/2}$ = 75.1 eV). Peaks were charge referenced to C-C band in the carbon 1s spectra at 284.4 eV. Measurements were made at a takeoff angle of 45° with respect to the sample surface plane. This resulted in a typical sampling depth of 3-6 nm (95% of the signal originated from this depth or shallower). Quantification was done using instrumental relative sensitivity factors (RSFs) that account for the X-ray cross section and inelastic mean free path of the electrons. These were derived from fresh exfoliated reference $WS_2$ and $MoS_2$ samples from HQ Graphene.

The $rGO/WS_2$ and $rGO/MoS_2$ films with similar dendritic-like morphology were formed, as shown in the SEM images (Figure S4a, d). EDS mappings in Figure S4b-c indicate that tungsten and sulfur were mostly detected on the dendritic-branches with the $WS_2$ stoichiometry, as well as $MoS_2$ (Figure S4e-f).

We prepared four different samples from precursor solution with different GO and ATTT weight ratios to investigate the role of GO working as surfactant in the formation of $WS_2$. Raman was used to study the structure of the synthetic samples (Figure S5a). It has been reported that under 514.5 nm laser excitation, the longitudinal acoustic mode *2LA(M)* (*ca.* 352 cm$^{-1}$) of $WS_2$ monolayers has a remarkable increase in intensity due to an active double resonance process, which causes that the intensity ratio of *2LA(M)* to $A_{1g}$ ($I_{2LA}/I_{A1g}$) almost reach 2.2 in monolayer as shown in Figure S5b.[1] As layer number increases to 3 layers, this intensity ratio ($I_{2LA}/I_{A1g}$) decreases to *ca.* 0.7.[1] It can be seen that the intensity ratio of $I_{2LA}/I_{A1g}$ increases with the reduction of ATTT in the precursor solution (see Figure S5c). The highest intensity ratio of $I_{2LA}/I_{A1g}$ obtained is *ca.* 1.33 when ATTT amount is the lowest (ATTT: GO=1:2). When ATTT amount increases 10 times (ATTT: GO=5:1), the intensity ratio of $I_{2LA}/I_{A1g}$ is *ca.* 1.00 (see Figure S5a, c). In general, this ratio ($I_{2LA}/I_{A1g}$) is always below 1.00 for $WS_2$ with a layer number exceeding 3 (see Figure S5b).[1] The obtained intensity ratio $I_{2LA}/I_{A1g}$ ranging from 1.00 to 1.33 suggests the presence of monolayers and bilayers of $WS_2$, exposing more catalytic active surface area when compared to the bulk structure. However, since low ATTT amounts result in smaller $WS_2$ coverage, the sample prepared with ATTT: GO=1:1 has the optimal $WS_2$ coverage and the optimal $I_{2LA}/I_{A1g}$ ratio, which is about 1.27. This material was used for further structural characterization and catalytic HER.

To further investigate the distribution of $WS_2$ layer numbers within the film, a Raman mapping using the 514.5 nm laser excitation was recorded (Figure S6). The optical image shown in Figure S6a displays the similar dendritic-like morphology as the SEM image (Figure S4a). It is noteworthy that the intensity ratio of *2LA(M)* to $A_{1g}$ is equal to or higher than 1 across the film, especially the $WS_2$ accumulated dendritic-branches on which the ratio is *ca.* 1.4 (Figure S6d), indicating that fewer layers $WS_2$ was formed across the film.

After transferring the films onto TEM grids, vertically aligned heterostructures of $rGO/WS_2$ was found from the dendritic area (Figure S7a). Discontinuous laminar structures constructed from mono- (Figure S7b) and few-layered (Figure S7c) $WS_2$ with rGO layers were found in the vertically aligned $rGO/WS_2$ by HRTEM. The intensity line profiles indicate that few-layer $WS_2$ with ~0.68 nm interlayer spacing is sandwiched with rGO layers exhibiting ~0.33 nm interlayer spacing, as well as monolayer $WS_2$ within the rGO layers.

High-resolution transmission electron microscopy (HRTEM) observations were performed in a JEM-2010F (JEOL) equipped with a field emission electron source and an ultrahigh resolution pole piece (Cs = 0.5 nm). An accelerating voltage at 200 kV was used for imaging.

The scanning transmission electron microscopy (STEM) observation was carried out by FEI Titan[3] G2 S/TEM operated at 80 keV and the high-angle annular dark field (HAADF) detector was used for *Z*-contrast imaging.

**Theoretical DFT Calculations (thermodynamics and kinetics): The Chemical Reaction Mechanism on $W_xMo_{1-x}S_2$.**

The Tafel slope $b$ can be expressed by the mathematical expression $b = 2.3RT/nF$; where $R$ is the universal gas constant (8.314458 J K$^{-1}$ mol$^{-1}$), $F$ is the Faraday constant (96485.33289), and $n$ is the difference between in the number of electrons between the ground state and the transition state. In the present HER reaction, $n = 1$ since the ground state has been shifted to the negatively charged structure as shown in Figure S12. The turn over frequency (TOF) of the hybrid $W_xMo_{1-x}S_2$ materials gradually increases due to presence of both W and Mo atoms, and the activation barrier decreases as shown in the Table S2 and Table S3. Our DFT calculations reveal that the $MoS_2$ has the lowest TOF among all the systems studied in this work, while $W_{0.4}Mo_{0.6}S_2$ has the highest.

**Electronic Properties Calculations (Band structures and density of states (DOSs)): the rGO/$W_{0.4}Mo_{0.6}S_2$ heterostructure**.

The electronic properties calculations (e.g. band and DOSs) were carried out to get a fundamental understanding of the synergistic effects leading to the electrocatalytic activity of the heterostructures formed by rGO with pristine ($MoS_2$, $WS_2$) as well as alloyed materials ($W_xMo_{1-x}S_2$, where x=0.1 − 0.9) as shown in Figure S11 and Figure 5a. We found that the pristine monolayer $MoS_2$ and $WS_2$ are pure semiconductors with high band gap ~2.5eV calculated at B3LYP-D2 level of theory. The calculated lattice constants (*a* and *b*) are 2.452 Å, 3.196 Å and 3.176 Å for graphene, $MoS_2$ and $WS_2$ respectively. We considered a 4x4 supercell to minimize the lattice mismatch between rGO and the $W_xMo_{1-x}S_2$ layer. Visualization were performed using VESTA[2] and additional analyses were performed using our in-house codes and scripts.

**Supplementary Figures and Tables**

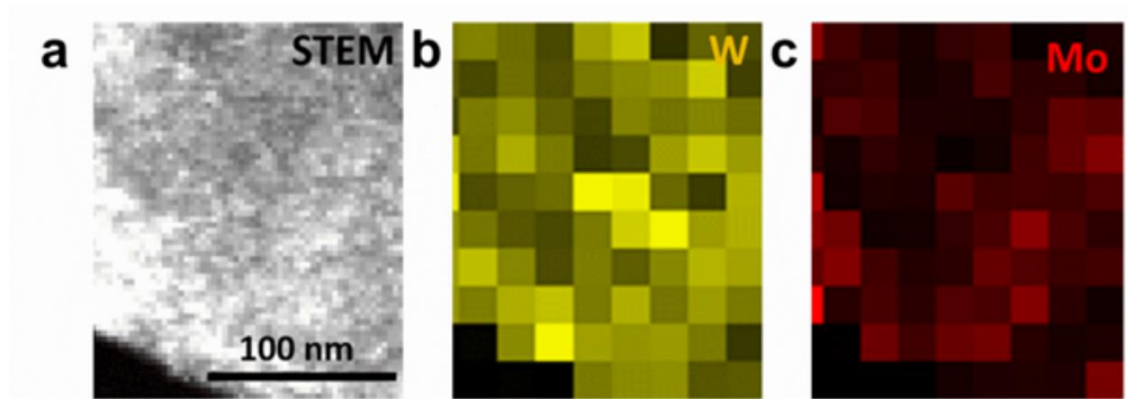

**Figure S1.** (a) STEM image of the rGO/WMoS$_2$ sample transferred to TEM grid. Elemental mapping by energy dispersive X-ray spectroscopy (EDS) of (b) tungsten (W(M) line); (c) molybdenum (Mo(K) line).

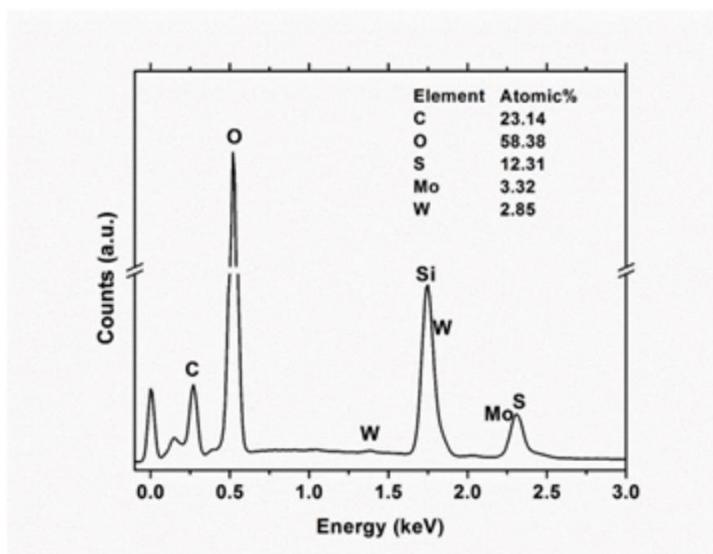

**Figure S2.** EDS spectrum obtained from the rGO/W$_x$Mo$_{1-x}$S$_2$ in Figure 1b, and the atomic ratio between W to Mo is 1:1.16 (x=0.46).

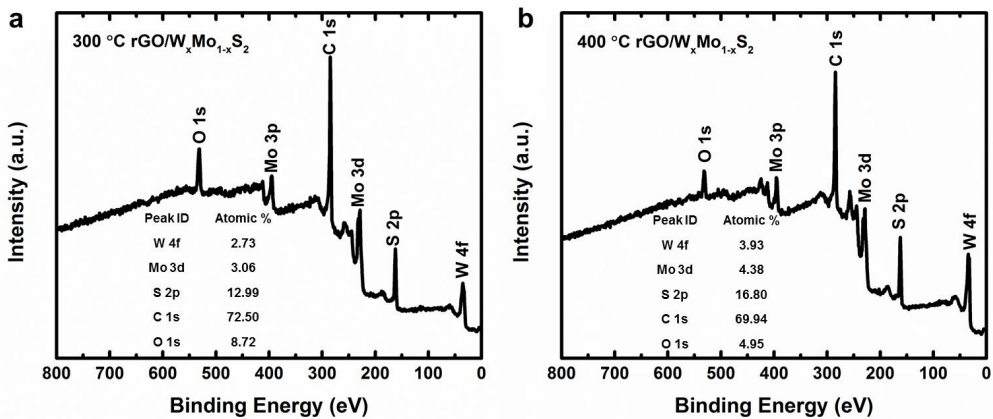

**Figure S3.** XPS survey spectra of rGO/$W_xMo_{1-x}S_2$ annealed at (a) 300 °C and (b) 400 °C; the inset tables summarized the elemental composition. Based on the elemental composition, the x values in rGO/$W_xMo_{1-x}S_2$ are 0.471 and 0.472 in the sample annealed at 300 and 400 °C, respectively. And the atomic ratio of S:(Mo+W) values are 2.24 and 2.02 in the sample annealed at 300 and 400 °C, respectively.

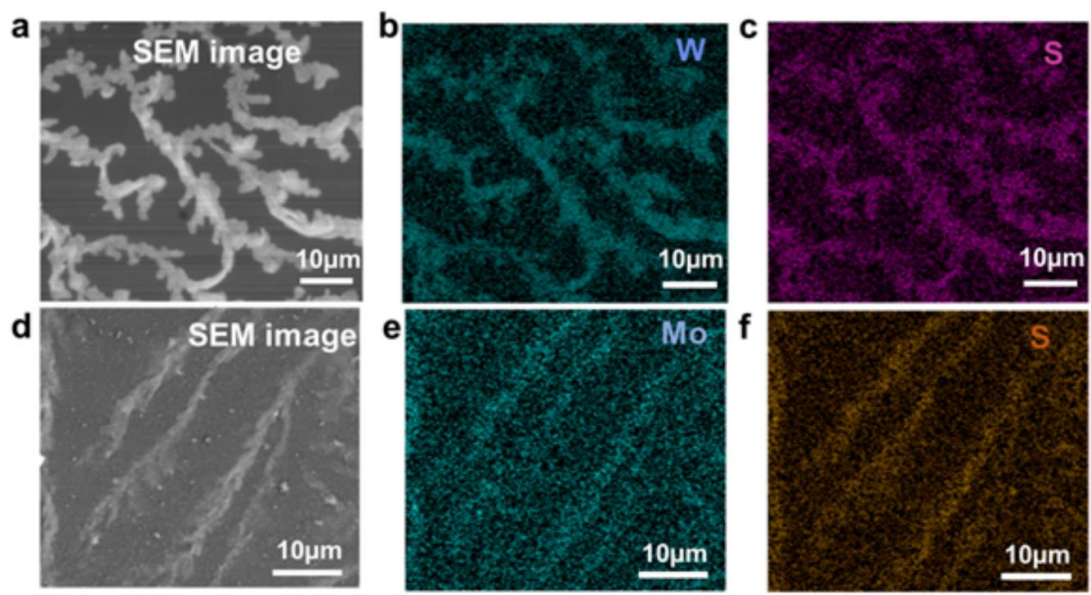

**Figure S4.** (a) SEM image of the rGO/WS$_2$ sample on a Si/SiO$_2$ substrate. EDS elemental mappings of: (b) tungsten (W-M line); (c) sulfur (S-K line); (d) SEM image of the rGO/MoS$_2$ sample on a Si/SiO$_2$ substrate; EDS mappings of (e) molybdenum (Mo-L line); (f) sulfur (S-K line).

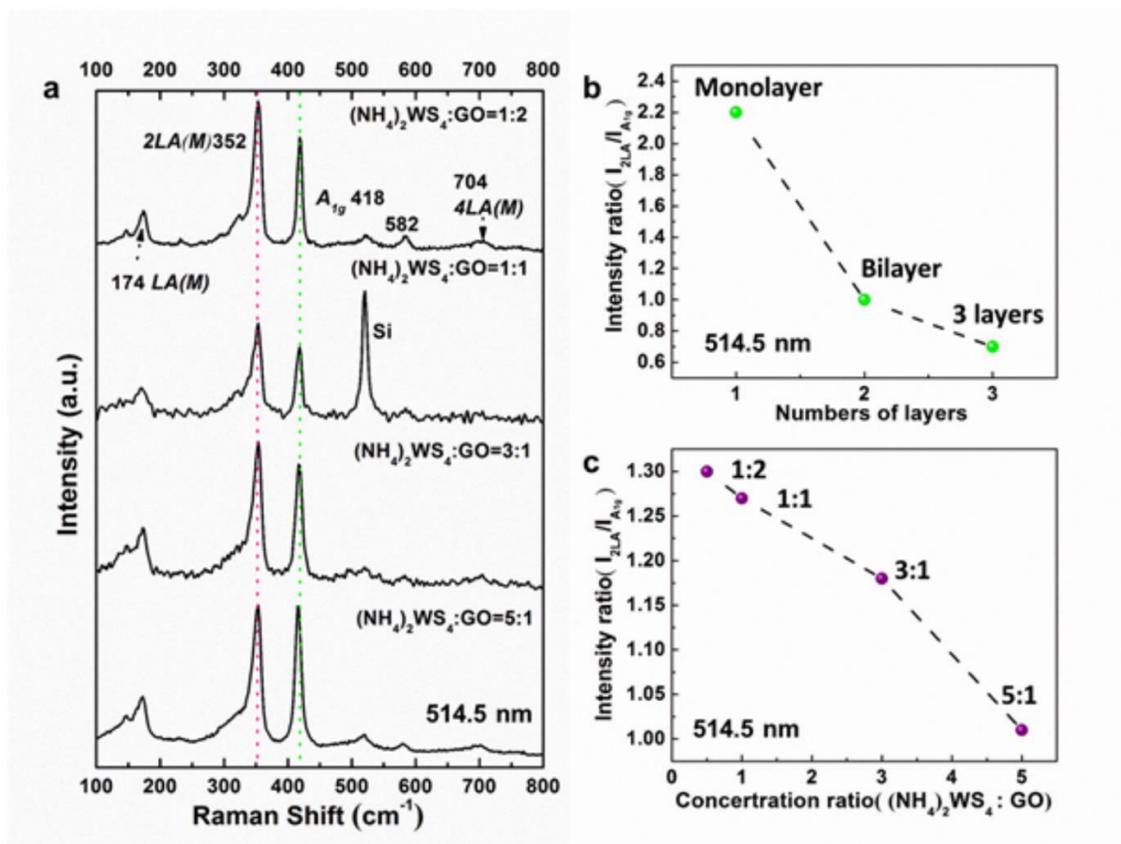

**Figure S5.** (a) Raman spectra of rGO/WS$_2$ samples with different ratios of (NH$_4$)$_2$WS$_4$/GO used in the precursor; (b) The intensity ratios of $I_{2LA}/I_{A1g}$ of WS$_2$ excited by a 514.5 nm laser with the corresponding number of layers,[1] and (c) The intensity ratios of $I_{2LA}/I_{A1g}$ in rGO/WS$_2$ samples with different ATTT/GO precursor amounts (excitation 514.5 nm laser).

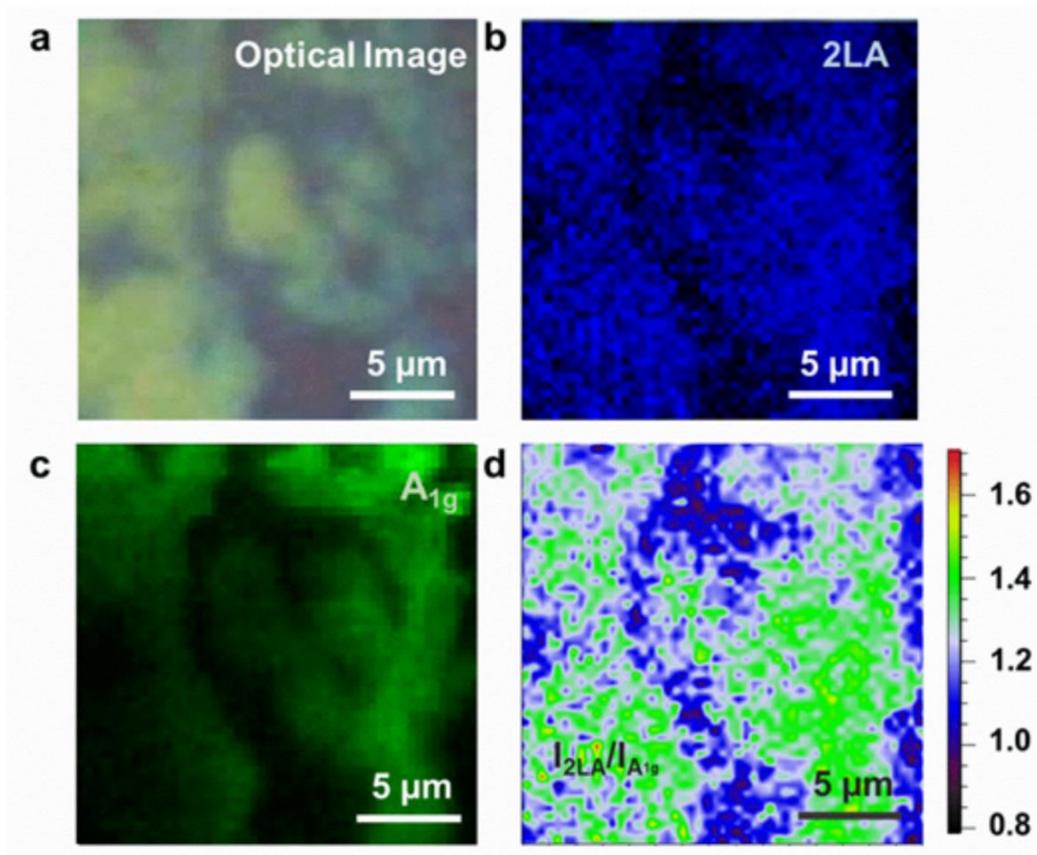

**Figure S6.** Raman mapping characterization of the rGO/WS$_2$ film. (a) Optical image. (b) *2LA(M)* and (c) *A$_{1g}$* intensity maps. (d) Intensity ratio map of *2LA(M)* to *A$_{1g}$* modes.

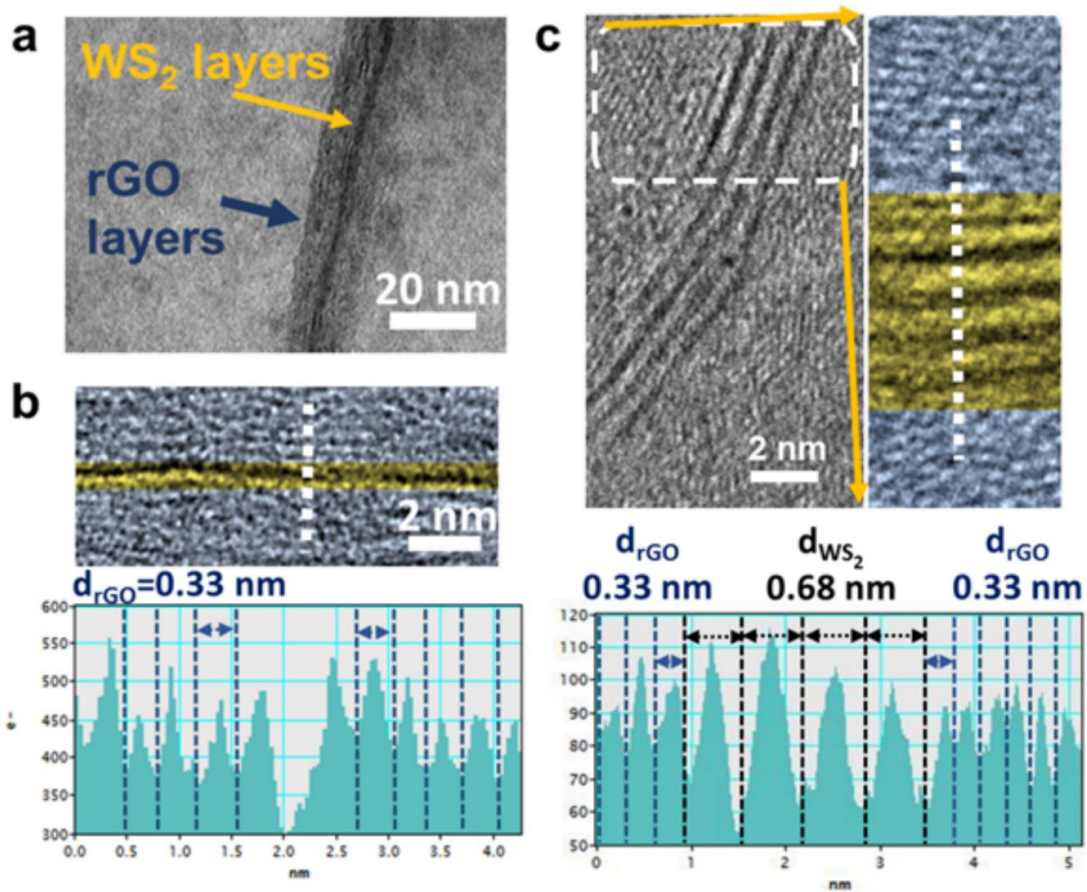

**Figure S7.** (a) TEM image of dendritic-branches of the rGO/WS$_2$ film; HRTEM image of (b) WS$_2$ monolayer covered by rGO layer (top panel), and the intensity line profile (bottom panel) of the white dashed line in HRTEM image; and (c) few-layers (top panel) covered by rGO layer, and the intensity line profile (bottom panel) of the white dashed line in HRTEM image. WS$_2$ layers are marked in yellow and carbon are highlighted in blue.

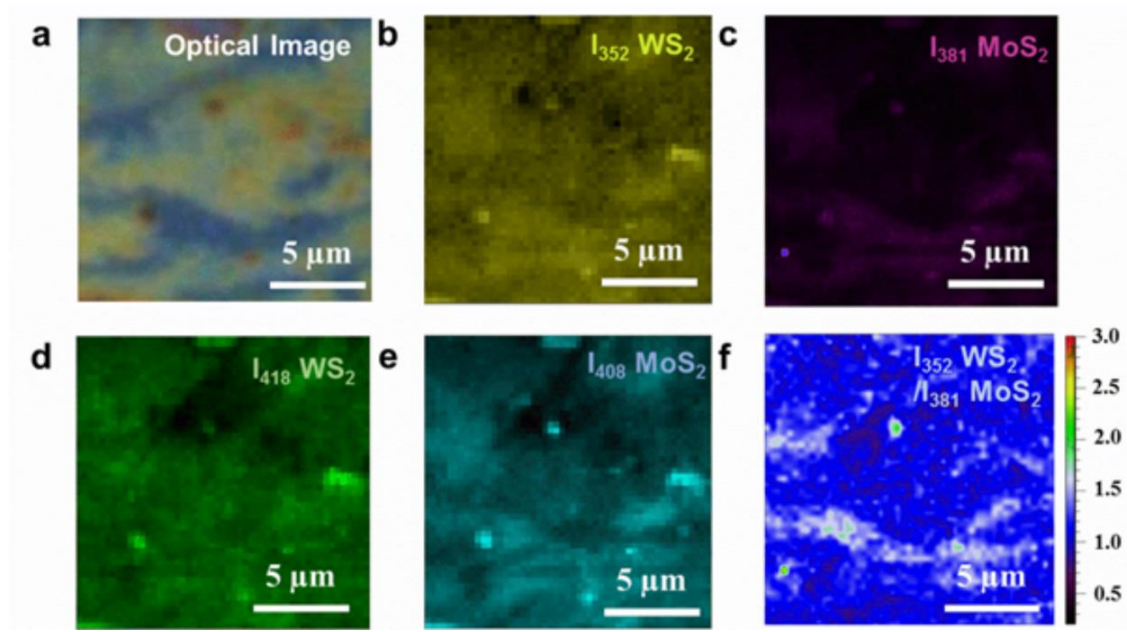

**Figure S8.** (a) Optical image of rGO/W$_x$Mo$_{1-x}$S$_2$; Raman intensity maps of (b) WS$_2$ *2LA* peak
(352 cm$^{-1}$), (c) MoS$_2$ *E$_{2g}$* peak (381 cm$^{-1}$), (d) WS$_2$ *A$_{1g}$* peak (418 cm$^{-1}$), and (e) MoS$_2$ *A$_{1g}$* peak
(408 cm$^{-1}$); (f) Intensity ratio map of WS$_2$ *2LA* over MoS$_2$ *E$_{2g}$*.

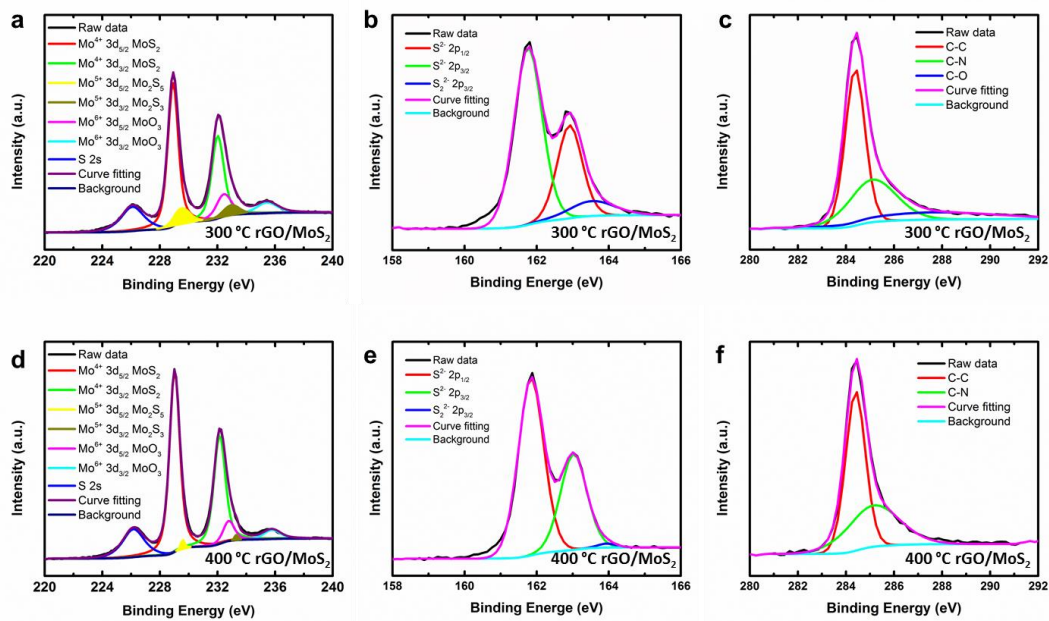

**Figure S9.** High-resolution XPS analysis of (a) Mo 3d, (b) S 2p, and (c) C 1s peaks of rGO/MoS$_2$ annealed at 300 °C; and the high-resolution XPS analysis of (d) Mo 3d, (e) S 2p, and (f) C 1s peaks of rGO/MoS$_2$ annealed at 400 °C. The Mo$^{5+}$ 3d$_{5/2}$ and Mo$^{5+}$ 3d$_{3/2}$ peaks are filled with yellow and dark yellow color, respectively.

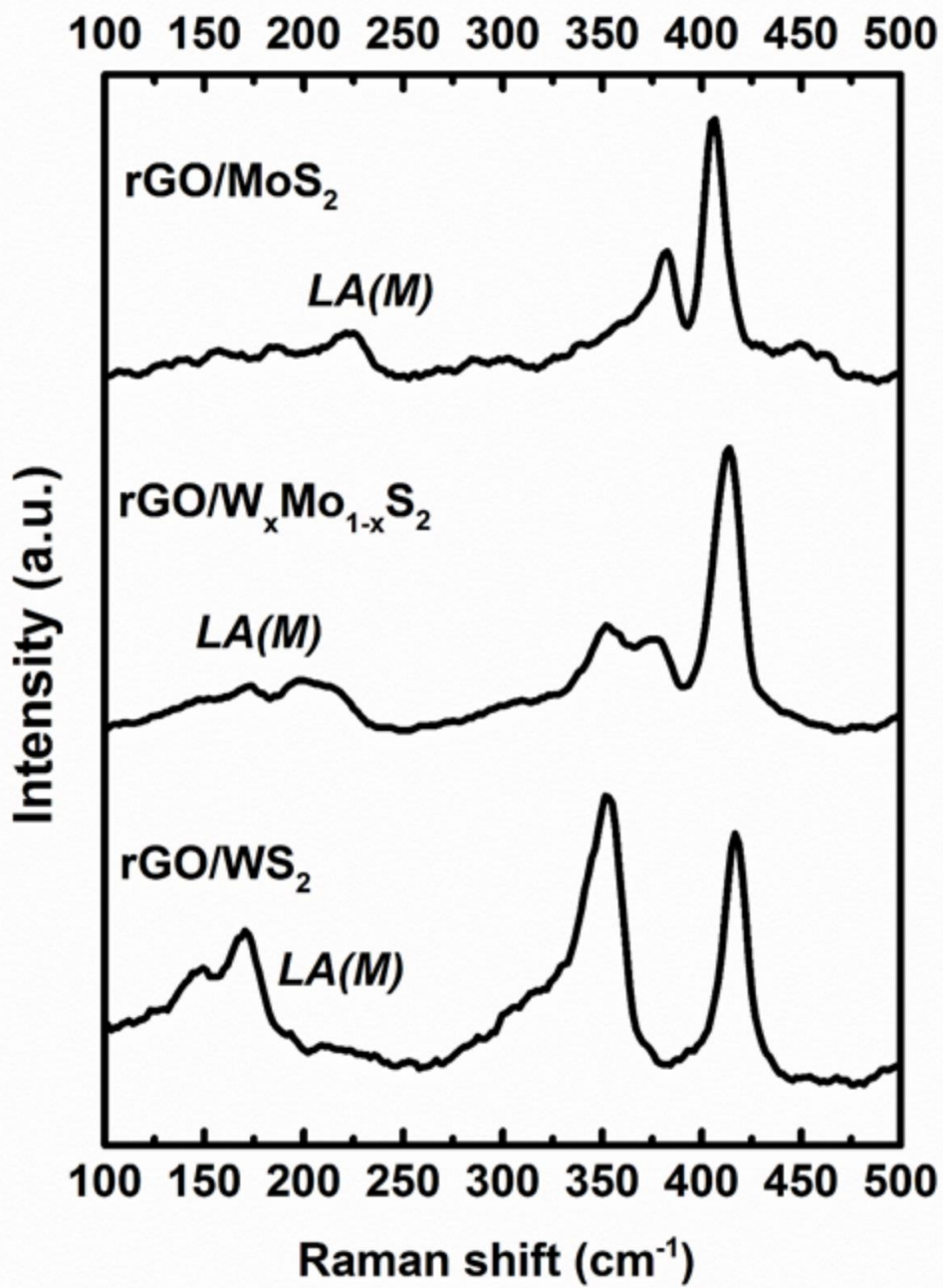

**Figure S10.** Raman spectra of rGO/WS$_2$, rGO/W$_x$Mo$_{1-x}$S$_2$, and rGO/MoS$_2$ annealed at 300 °C.

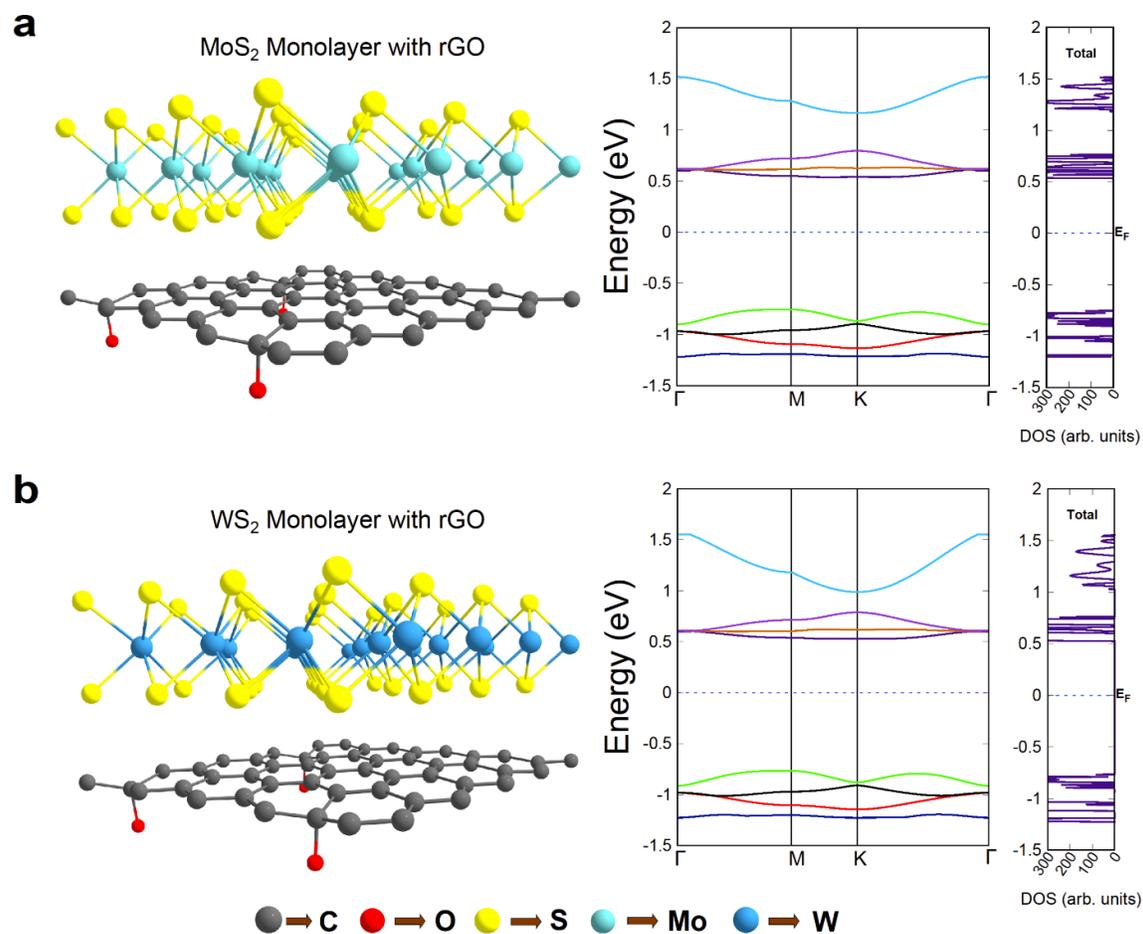

**Figure S11.** Electronic structures, band structures and density of states (DOSs) of the heterostructures formed between rGO and (a) MoS$_2$ and (b) WS$_2$.

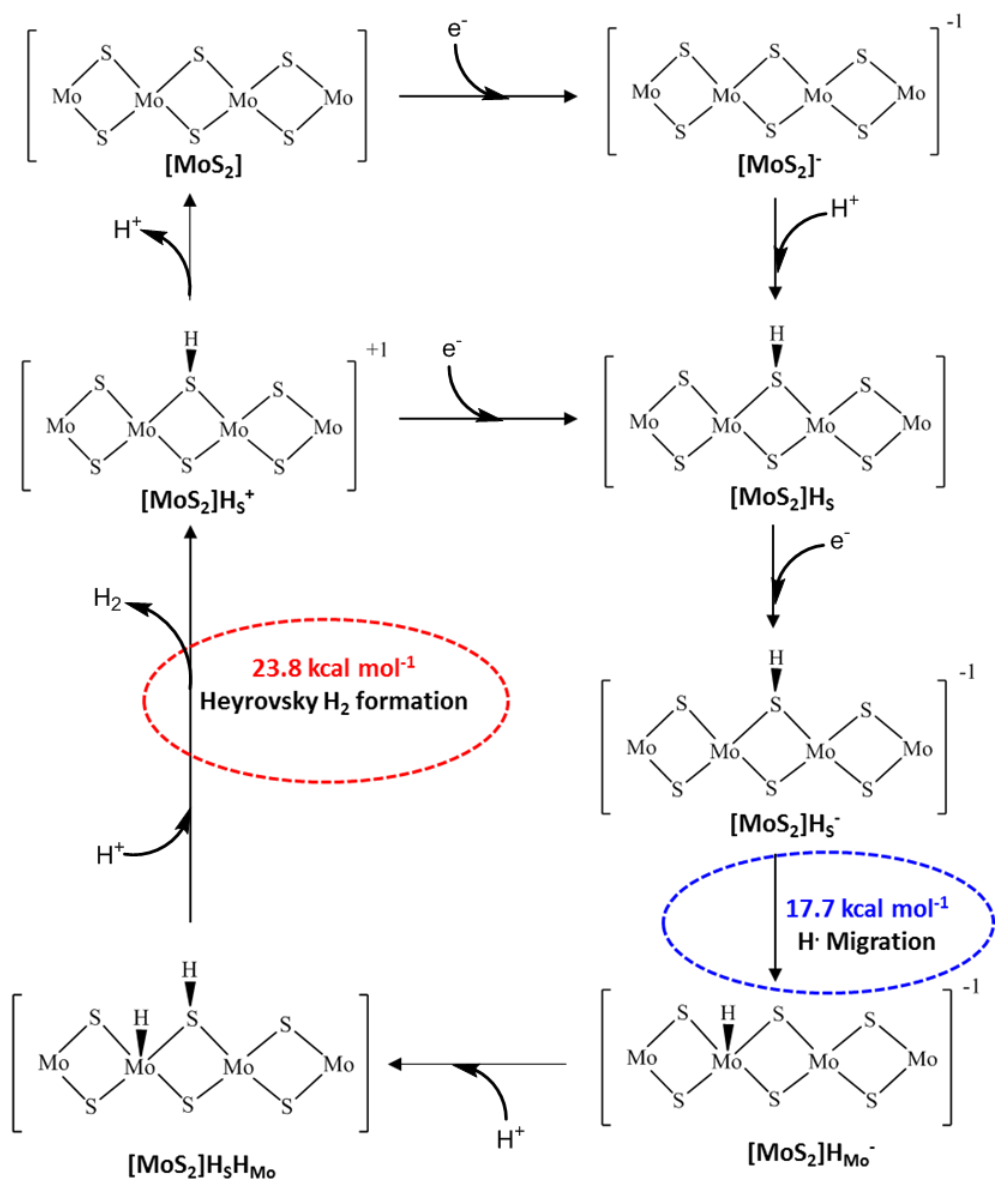

**Figure S12.** Overall chemical reaction mechanism for $H_2$ evolution on $MoS_2$ materials using the DFT method with the M06-L functional. M06-L has shown to give accurate energies for reaction barriers involving TMs.[3] The H· migration reaction barrier (Volmer reaction mechanism, highlighted in a dotted blue circle) and the $H_2$ formation (Heyrovsky reaction mechanism, highlighted in a dotted red circle) on $MoS_2$ materials are 17.7 kcal mol$^{-1}$ and 23.8 kcal mol$^{-1}$, respectively.

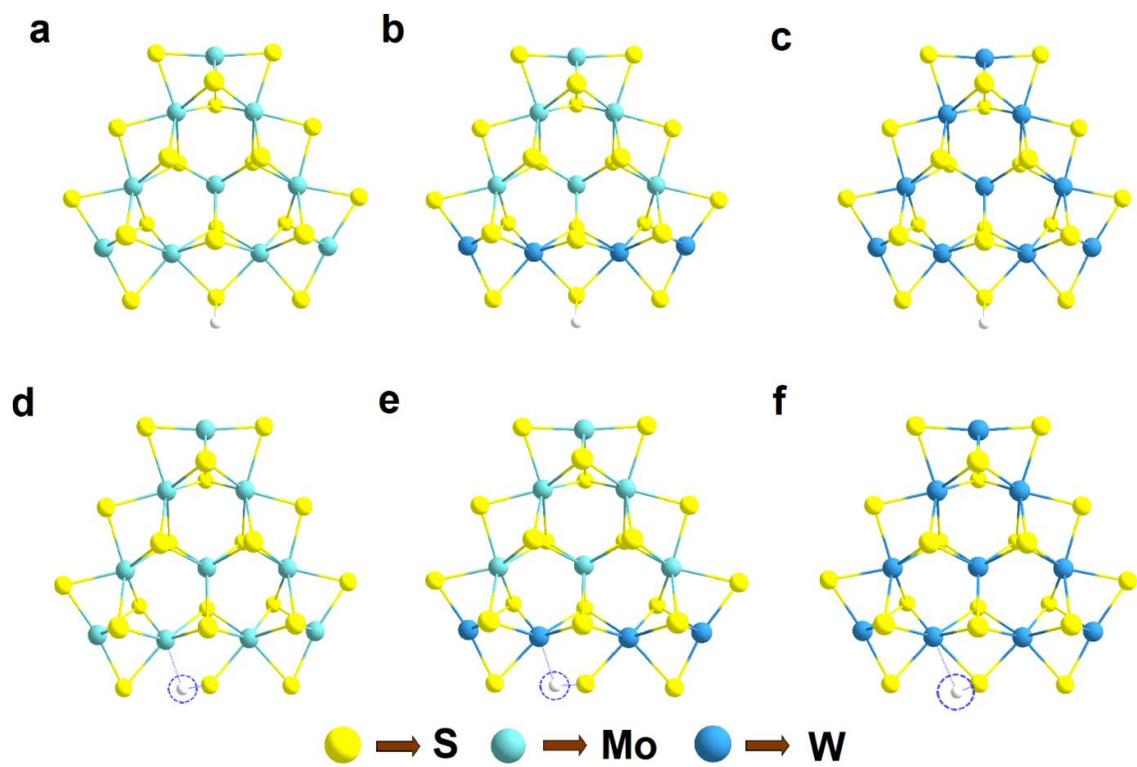

**Figure S13.** The optimized geometries of (a) $MoS_2H_S$, (b) $W_xMo_{1-x}S_2H_S$ (here x=0.4), and (c) $WS_2H_S$ materials considered as a model system in the DFT calculations. The transition states (TSs) occurred in the H· migration reaction of these materials are presented in d, e and f, and the position of the H in the TSs is highlighted by a dotted blue circle.

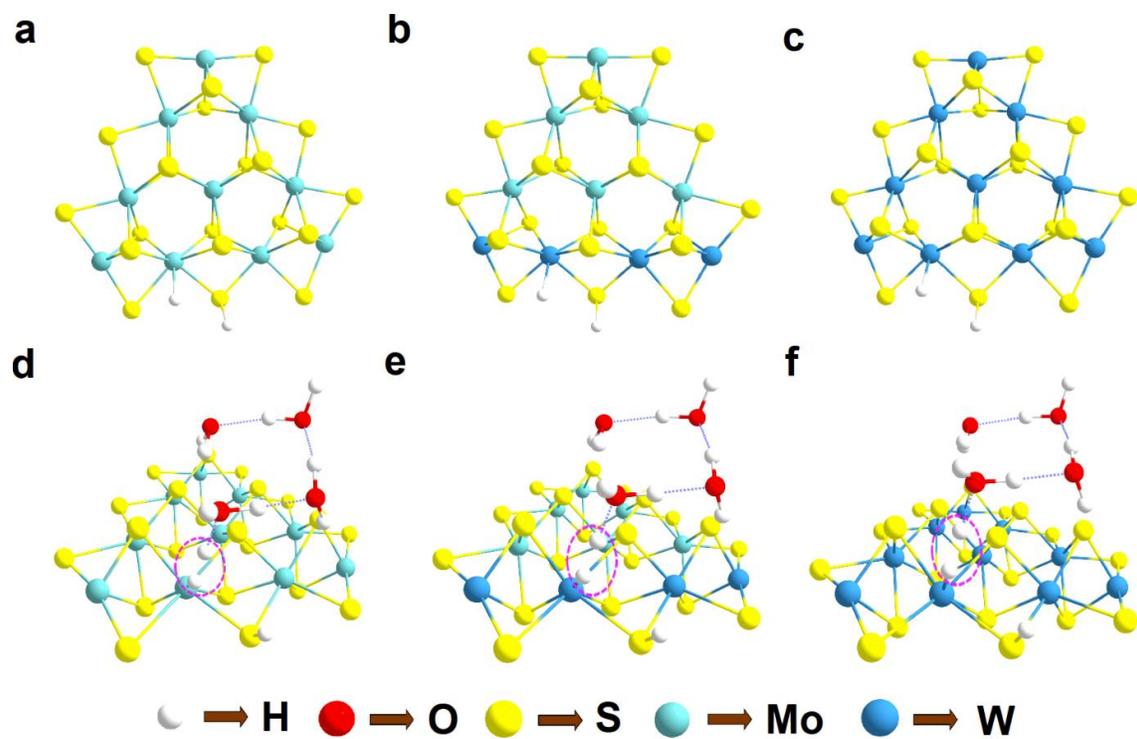

**Figure S14.** The optimized geometries of (a) $MoS_2H_SH_{Mo}$, (b) $W_xMo_{1-x}S_2H_SH_W$ (here x=0.4), and (c) $WS_2H_SH_W$ materials considered in the DFT calculations. The transition states (TSs) occurred in the Heyrovsky reaction of these materials, TS1, TS2 and TS3 are presented in d, e and f, and the position of the $H_2$ in the TSs is highlighted by a dotted pink circle.

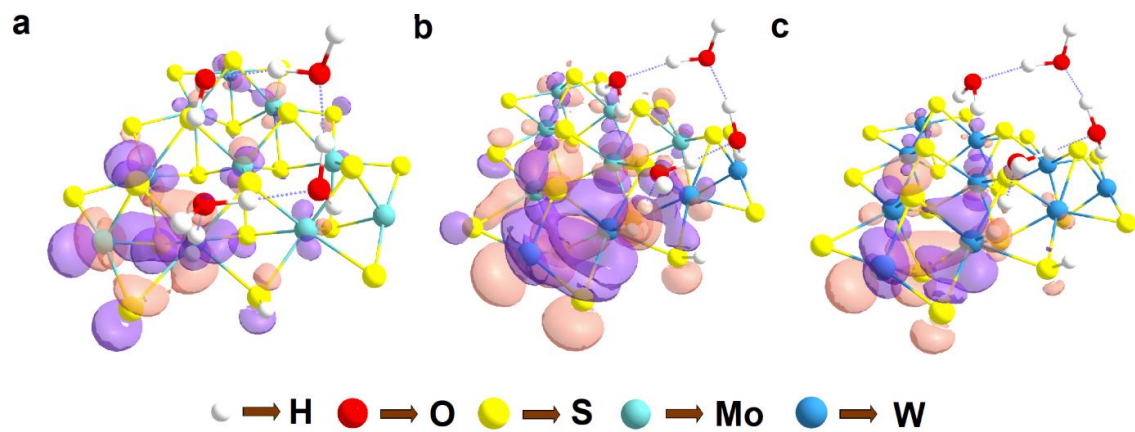

**Figure S15.** Top view of the highest occupied molecular orbital (HOMO) of the transition state structures found in the Heyrovsky reaction mechanism in the presence of four explicit water cluster for: (a) $MoS_2$ (TS1), (b) $W_{0.4}Mo_{0.6}S_2$ (TS2), and (c) $WS_2$ (TS3). The molecular orbitals involved in the Heyrovsky reaction and activation barrier in HER are highlighted by a dotted pink circle.

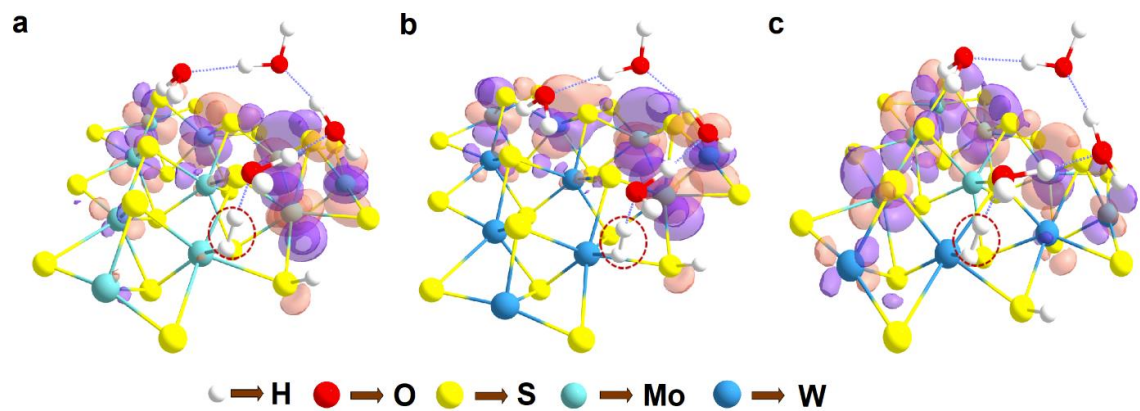

**Figure S16.** Lowest unoccupied molecular orbital (LUMO) of the transition state structures found in the Heyrovsky reaction mechanism in the presence of four explicit water cluster for: (a) $MoS_2$ (TS1), (b) $W_{0.4}Mo_{0.6}S_2$ (TS2), and (c) $WS_2$ (TS3). The position of the molecular $H_2$ involved in Heyrovsky reaction are highlighted by a dotted red circle.

**Table S1.** Electrochemical impedance spectroscopy (EIS) fitting and double layer capacitance ($C_{dl}$) calculation results.

| Samples (This work) | $R_p$ $\Omega$ | CPE-T | CPE-P | $C_{dl}$ mF cm$^{-2}$ |
|---|---|---|---|---|
| rGO/W$_x$Mo$_{1-x}$S$_2$ (300 °C) | 6.65 | 0.0168 | 0.7903 | 23.39 |
| rGO/MoS$_2$ (300 °C) | 7.79 | 0.0122 | 0.8253 | 13.25 |
| rGO/WS$_2$ (300 °C) | 2.69 | 0.0278 | 0.8633 | 37.61 |
| rGO/W$_x$Mo$_{1-x}$S$_2$ (400 °C) | 5.73 | 0.0177 | 0.7531 | 17.05 |
| rGO/MoS$_2$ (400 °C) | 4.43 | 0.0252 | 0.7580 | 25.61 |
| rGO/WS$_2$ (400 °C) | 3.01 | 0.0510 | 0.7925 | 39.29 |

**Table S2.** H$^.$ migration reaction energy barriers (Volmer reaction mechanism) obtained for MoS$_2$, WS$_2$, and hybrid W$_x$Mo$_{1-x}$S$_2$ (x=0.1 − 0.9) materials.

| Compounds | Barrier in kcal mol$^{-1}$ (Gas Phase) | Barrier in kcal mol$^{-1}$ (Solvent, water) |
|---|---|---|
| MoS$_2$ | 11.9 | 17.7 |
| W$_{0.1}$Mo$_{0.9}$S$_2$ | 10.2 | 14.0 |
| W$_{0.2}$Mo$_{0.8}$S$_2$ | 09.3 | 13.9 |
| W$_{0.3}$Mo$_{0.7}$S$_2$ | 08.0 | 12.4 |
| W$_{0.4}$Mo$_{0.6}$S$_2$ | 06.8 | 11.9 |
| W$_{0.5}$Mo$_{0.5}$S$_2$ | 08.1 | 13.0 |
| W$_{0.6}$Mo$_{0.4}$S$_2$ | 09.6 | 14.9 |
| W$_{0.7}$Mo$_{0.3}$S$_2$ | 10.1 | 12.4 |
| **W$_{0.8}$Mo$_{0.2}$S$_2$** | 10.7 | 13.1 |
| **W$_{0.9}$Mo$_{0.1}$S$_2$** | 11.2 | 15.8 |
| **WS$_2$** | 12.4 | 18.1 |

**Table S3.** H$_2$ formation reaction energy barriers (Heyrovsky reaction mechanism) obtained for MoS$_2$, WS$_2$, and W$_x$Mo$_{1-x}$S$_2$ (x=0.1 – 0.9) with four explicit water molecules and Turnover Frequency (TOF). The unit of TOF is H$_2$ sec$^{-1}$ per edge Mo/W atom.

| Compounds | Barrier in kcal mol$^{-1}$ (Gas Phase) | Barrier in kcal mol$^{-1}$ (Solvent, water) | TOF in sec$^{-1}$ (Solvent, water) |
|---|---|---|---|
| **MoS$_2$** | 16.0 | 23.8 | 2.1 x 10$^{-5}$ |
| **W$_{0.1}$Mo$_{0.9}$S$_2$** | 13.8 | 20.1 | 1.2 x 10$^{-2}$ |
| **W$_{0.2}$Mo$_{0.8}$S$_2$** | 12.5 | 18.0 | 4.2 x 10$^{-1}$ |
| **W$_{0.3}$Mo$_{0.7}$S$_2$** | 11.9 | 15.3 | 3.5 x 10$^{1}$ |
| **W$_{0.4}$Mo$_{0.6}$S$_2$** | 11.5 | 13.3 | 1.1 x 10$^{3}$ |
| **W$_{0.5}$Mo$_{0.5}$S$_2$** | 12.2 | 14.0 | 3.3 x 10$^{2}$ |
| **W$_{0.6}$Mo$_{0.4}$S$_2$** | 13.0 | 15.1 | 5.1 x 10$^{1}$ |
| **W$_{0.7}$Mo$_{0.3}$S$_2$** | 13.3 | 17.1 | 1.9 x 10$^{0}$ |
| **W$_{0.8}$Mo$_{0.2}$S$_2$** | 13.5 | 18.4 | 2.1 x 10$^{-1}$ |
| **W$_{0.9}$Mo$_{0.1}$S$_2$** | 14.0 | 19.8 | 1.7 x 10$^{-2}$ |
| **WS$_2$** | 14.5 | 21.3 | 1.5 x 10$^{-3}$ |